\documentclass[aps,prb,showpacs,superscriptaddress,twocolumn,amsmath,amssymb,floatfix]{revtex4}
\pdfoutput=1
\usepackage{graphicx}
\usepackage{bm}
\usepackage{fancybox}

\begin{document}

\title{Density-Functional Theory of Graphene Sheets}

\author{Marco Polini}
\email{m.polini@sns.it}
\affiliation{NEST-CNR-INFM and Scuola Normale Superiore, I-56126 Pisa, Italy}
\author{Andrea Tomadin}
\affiliation{NEST-CNR-INFM and Scuola Normale Superiore, I-56126 Pisa, Italy}
\author{Reza Asgari}
\affiliation{School of Physics, Institute for Studies in Theoretical Physics and
Mathematics, 19395-5531 Tehran, Iran}
\author{A.H. MacDonald}
\affiliation{Department of Physics, The University of Texas at Austin, Austin Texas 78712}

\begin{abstract}
We outline a Kohn-Sham-Dirac density-functional-theory (DFT) scheme for graphene sheets that 
treats slowly-varying inhomogeneous external potentials and electron-electron interactions
on an equal footing. The theory is able to account for the the unusual property that 
the exchange-correlation contribution to chemical potential increases with 
carrier density in graphene.  Consequences of this property, and advantages and disadvantages of using 
the DFT approach to describe it, are discussed. 
The approach is illustrated by solving the Kohn-Sham-Dirac equations self-consistently for a model 
random potential describing charged point-like impurities located close to the graphene plane. The influence of electron-electron interactions on these non-linear screening calculations is discussed at length, in the light of recent experiments reporting evidence for the presence of electron-hole puddles in nearly-neutral graphene sheets. 
\end{abstract}

\date{\today}

\maketitle

\section{Introduction}  

Graphene is a newly realized two-dimensional (2D) electron system~\cite{reviews,PT} 
which has engendered a great deal of interest because of the new physics which it exhibits and because of 
its potential as a new material for electronic technology.  The agent responsible for many of 
the interesting electronic properties of graphene sheets is the non-Bravais honeycomb-lattice
arrangement of carbon atoms, which leads to a gapless semiconductor
with valence and conduction $\pi$-bands.
States near the Fermi energy of a graphene sheet are described by a massless Dirac equation which has 
chiral band states in which the honeycomb-sublattice pseudospin is aligned either parallel to 
or opposite to the envelope function momentum.  The Dirac-like wave equation 
leads to both unusual electron-electron interaction effects and to unusual response to external potentials. 
 
Many new ideas that are now being explored in graphene electronics are still based on idealized models which neglect disorder and electron-electron interactions, and as a consequence many of these may ultimately require qualitative and quantitative revision as 
our understanding of this material improves. In this paper we outline one approach, a Kohn-Sham-Dirac DFT scheme,
which can be used for more realistic modelling of graphene sheets, including both disorder and electron-electron interactions.  

Because of band chirality, the role of electron-electron interactions in graphene sheets differs in some 
essential ways~\cite{barlas_prl_2007,ourdgastheory,dassarmadgastheory} 
from the role which it plays in an ordinary 2D electron gas.
One important difference is that the contribution of exchange and correlation to the chemical potential is 
an increasing rather than a decreasing function of carrier-density.  As we discuss later, this property 
implies that exchange and correlation increases the effectiveness of screening, in contrast to the usual case in which exchange and correlation weakens screening. This unusual property follows from the difference in sublattice pseudospin 
chirality between the Dirac model's negative energy valence band states and its conduction band states~\cite{barlas_prl_2007,ourdgastheory}, and in a uniform graphene system is readily accounted for by many-body perturbation theory.  The principle merit of the DFT theory we describe is that it allows this physics to be accounted for in graphene sheets in which the carrier density is non-uniform either by design, as in p-n junction systems~\cite{pnjunctionexpts}, or as a result of unintended disorder sources. 

A related and complementary DFT method has recently been used by Rossi and Das Sarma~\cite{enrico_condmat} to study the ground-state density profile of massless Dirac fermions in the presence of randomly-distributed charged impurities. Their method differs from ours in two main respects: the authors of Ref.~\onlinecite{enrico_condmat} have (i) approximated the kinetic energy functional of non-interacting massless Dirac fermions by means of a local-density approximation (LDA) whereas in the present work the kinetic energy functional is treated exactly {\it via} the Kohn-Sham mapping (see Sect.~\ref{sect:theory} below); and (ii) neglected correlation effects, which, as it will be clear in Sect.~\ref{sect:c}, partly compensate the enhanced screening due to exchange and Dirac-equation chirality.  Inhomogeneous graphene systems have also been studied using the Thomas-Fermi approximation (LDA for the kinetic energy only) by Fogler and collaborators~\cite{fogler}  .

Our paper is organized as follows.  In Section~\ref{sect:theory} we outline the version of DFT which is appropriate for non-uniform carrier density graphene sheets with static external potentials that are smooth enough to permit neglect of inter-valley scattering. Many-body effects enter this theory via an LDA exchange-correlation potential
with a density-dependence precisely opposite to the one familiar from ordinary LDA-DFT theory 
applied to parabolic-band inhomogeneous electron liquids.  In Section~\ref{sect:pw} we outline the procedure we have used to solve the theory's Dirac-like Kohn-Sham equations.  In Section~\ref{sect:nls} we discuss results obtained by solving the Kohn-Sham equations self-consistently for an illustrative random potential model, highlighting some strengths and weaknesses of this approach to many-body physics in inhomogeneous graphene sheets.
In Section~\ref{sect:conclusions} we briefly mention other problems to which the theory outlined in this paper could be successfully applied, comment on the relationship between our DFT approach and {\it ab~initio} DFT applied to graphene, and summarize our main conclusions.

\section{Massless Dirac Model DFT}  
\label{sect:theory}

We consider a system of 2D massless Dirac fermions which are subjected to a 
time-independent scalar external potential $V_{\rm ext}({\bm r})$.  
This model applies to graphene sheets when the 
external potential varies slowly
on the lattice-constant length scale.  In this limit the external potential
will couple identically to the two sublattices, and hence be a pseudospin scalar,
and have negligible inter-valley scattering, justifying an envelope function approach~\cite{cardona}  
which promotes the perfect crystal Dirac bands to envelope function Dirac operators. 
To account for electron-electron interactions in graphene sheets, the 
ultrarelativistic massless-Dirac particles must interact via instantaneous non-relativistic
Coulomb interactions.  The juxtaposition of an ultrarelativistic free-fermion term and a 
non-relativistic interaction term in the Hamiltonian of a graphene sheet
leads to a new-type of many-body problem.  

DFT~\cite{dft,dft_reviews, Giuliani_and_Vignale} is a practical 
approach to many-body physics which recognizes the impossibility of achieving exact results and seeks practical solutions with adequate accuracy.
Following a familiar line of argument~\cite{dft,dft_reviews,Giuliani_and_Vignale} which we do not 
reproduce here, many-body exchange-correlation effects can be taken into account in the 
graphene many-body problem with the same formal justifications and the 
same types of approximation schemes as in standard non-relativistic 
DFT~\cite{dft,dft_reviews,Giuliani_and_Vignale}. 
The end result in the case of present interest is that 
ground state charge densities and energies are determined by solving a 
time-independent Kohn-Sham-Dirac equation for a sublattice-pseudospin  
spinor $\Phi_\lambda({\bm r})= (\varphi^{(A)}_{\lambda}({\bm r}),\varphi^{(B)}_{\lambda}({\bm r}))^{\rm T}$,
\begin{equation}\label{eq:ksd}
\left[v{\bm \sigma}\cdot {\bm p}+{\mathbb I}_{\sigma}V_{\rm KS}({\bm r})\right]\Phi_\lambda({\bm r})=\varepsilon_{\lambda}\Phi_\lambda({\bm r})~.
\end{equation}
Here $v \sim 10^{6}~{\rm m}/{\rm s}$ is the bare Fermi velocity, ${\bm p}=-i\hbar \nabla_{\bm r}$, ${\bm \sigma}$ is a 2D vector constructed with the $2 \times 2$ Pauli matrices $\sigma_1$ and $\sigma_2$ acting in pseudospin space, 
${\mathbb I}_{\sigma}$ is the $2 \times 2$ identity matrix in pseudospin space, and $V_{\rm KS}({\bm r})=V_{\rm ext}({\bm r})+\Delta V_{\rm H}({\bm r})+V_{\rm xc}({\bm r})$ is the Kohn-Sham (KS) potential, which is a functional of the ground-state density $n({\bm r})$.  
The ground-state density is obtained as a sum over occupied 
Kohn-Sham-Dirac spinors $\Phi_\lambda({\bm r})$:
\begin{eqnarray}\label{eq:density}
n({\bm r}) & = & 4\sum_{\lambda ({\rm occ})}
\Phi^\dagger_{\lambda}({\bm r})\Phi_{\lambda}({\bm r})\nonumber\\
&\equiv&
4\sum_{\lambda ({\rm occ})}[|\varphi^{(A)}_{\lambda}({\bm r})|^2+|\varphi^{(B)}_{\lambda}({\bm r})|^2]~,
\end{eqnarray}
where the factor $4$ is due to valley and spin degeneracies and $\{\varphi^{(\sigma)}_{\lambda}({\bm r}), \sigma=A,B\}$ are the pseudospin (sublattice) components of the Kohn-Sham-Dirac spinor $\Phi_\lambda({\bm r})$. Equation (\ref{eq:density}) is a self-consistent closure relationship for the Kohn-Sham-Dirac equations (\ref{eq:ksd}), since the effective potential in Eq.~(\ref{eq:ksd}) is a functional of the ground-state density $n({\bm r})$. 
More explicit details on the construction of $n({\bm r})$ are given below.  This formalism is readily generalized to account for 
spin-polarization~\cite{spinpol}, or valley polarization~\cite{valleypol}, or both. 
A generalization of the present theory to situations in which graphene is subjected 
to an inhomogenous magnetic field (as in magnetically-defined graphene quantum dots~\cite{egger_prl_2007}) 
can also be envisioned along the lines of {\it e.g.} Ref.~\onlinecite{vignale_prl_1987}.

The KS potential $V_{\rm KS}({\bm r})$ in Eq.(~\ref{eq:ksd}) is the sum of external, Hartree, and exchange-correlation contributions.
The Hartree potential
\begin{equation}\label{eq:hartree}
\Delta V_{\rm H}({\bm r})=\int d^2{\bm r}'\frac{e^2}{\epsilon|{\bm r}-{\bm r}'|} \; \delta n({\bm r}')~,
\end{equation}
where $\epsilon$ is the average background dielectric constant 
($\epsilon=2.5$, for example, for graphene placed on ${\rm SiO}_2$ with the other side being exposed to air) 
and the quantity $\delta n({\bm r})=n({\bm r})-n_0$ is the density measured relative to that of a 
uniform neutral graphene sheet as specified more precisely below [see Eq.~(\ref{eq:renormalization})]. 

The third term in $V_{\rm KS}({\bm r})$, $V_{\rm xc}({\bm r})$, is the exchange-correlation potential, which is 
formally a functional of the ground-state density, but known only approximately.
In this work we employ the local-density approximation,
\begin{eqnarray}\label{eq:lda}
V_{\rm xc}({\bm r})&=&\left.v^{\rm hom}_{\rm xc}(n)\right|_{n \to n_{\rm c}({\bm r})}~,
\end{eqnarray}
where $v^{\rm hom}_{\rm xc}(n)$ is the reference exchange-correlation potential of a {\it uniform 2D liquid of 
massless Dirac fermions}~\cite{barlas_prl_2007,ourdgastheory} with carrier density $n$.  
$v^{\rm hom}_{\rm xc}(n)$ is related to the ground-state energy per excess carrier 
$\delta \varepsilon_{\rm xc}(n)$ by  
\begin{equation}
v^{\rm hom}_{\rm xc}(n)=\frac{\partial [n \delta\varepsilon_{\rm xc}(n)]}{\partial n}~.
\end{equation}
The carrier density $n_{\rm c}({\bm r})$ is the density relative to that of a uniform {\it neutral} graphene sheet and 
will be defined more precisely in Sect.~\ref{eq:construction}.  The expression used for 
$\delta \varepsilon_{\rm xc}(n)$ depends on the zero-of-energy, which is normally~\cite{barlas_prl_2007,ourdgastheory} chosen so that
$v^{\rm hom}_{\rm xc}(n=0)=0$.   

To apply the LDA-DFT formalism to graphene it is necessary to have convenient expressions for the excess exchange-correlation energy $\delta\varepsilon_{\rm xc}(n)$, which will be provided below in Sects.~\ref{sect:x} and~\ref{sect:c}. This quantity has been calculated at the Random Phase Approximation (RPA) level in Ref.~\onlinecite{barlas_prl_2007}.

\subsection{Exchange potential}
\label{sect:x}

Because the Coulomb energy and the Dirac band energy scale in the same 
way with length, we can write 
the first-order exchange contribution to $\delta\varepsilon_{\rm xc}(n)$ as
\begin{equation}\label{eq:ex_elegant}
\delta \varepsilon_{\rm x}(n)=\varepsilon_{\rm F} \alpha_{\rm gr} F(\Lambda)~.
\end{equation}
Here $\varepsilon_{\rm F}={\rm sgn}(n)~\hbar v k_{\rm F}$ is the Fermi energy, where 
$k_{\rm F}=(4 \pi |n|/g)^{1/2}$ is the Fermi wave vector corresponding to an electron (hole) density $n$ above (below) the neutrality point 
and $g = g_{\rm s} g_{\rm v}=4$ accounts for spin and valley degeneracy. The quantity $\alpha_{\rm gr}$ is defined by 
$\alpha_{\rm gr}=g\times e^2/(\epsilon \hbar v) \equiv g\times \alpha_{\rm ee}$, 
where $\alpha_{\rm ee}$ is graphene's fine structure constant. 
The ultraviolet cut-off $\Lambda$ in Eq.~(\ref{eq:ex_elegant}) is defined by $\Lambda = k_{\rm max}/k_{\rm F}$, where 
$k_{\rm max}$ 
should be assigned a value corresponding to the
wavevector range over which the continuum model describes graphene. 
For definiteness we take $k_{\rm max}$ to be such that
\begin{eqnarray}\label{eq:eta}
\pi k^2_{\rm max}=\eta \frac{(2\pi)^2}{{\cal A}_0}~,
\end{eqnarray}
where ${\cal A}_0=3\sqrt{3} a^2_0/2 \sim 0.052~{\rm nm}^2$ is the area of the unit cell in the honeycomb lattice. 
($a_0 \simeq 1.42$~\AA~is the Carbon-Carbon distance) and $\eta$ is a dimensionless number $\eta \in (0,1]$. 
The optimal value of $\eta$ would have to be determined by a lattice-model correlation energy calculation. 
From another point of view $\eta$, the Dirac velocity $v$, and the dielectric constant $\epsilon$ are 
coupled parameters of the Dirac model for graphene which should be fixed by comparison of the model's 
predictions with experiment.  For typical graphene-system densities, the dependence of the exchange-correlation potential on $\eta$ is weak enough 
that we can arbitrarily choose $\eta=1$ with some confidence.
Given a value of $\eta$, the dependence of $\Lambda$ on density is given by
\begin{eqnarray}
\Lambda(n) = \sqrt{g\eta}\frac{1}{\sqrt{|n| {\cal A}_0}}~.
\end{eqnarray}

The exchange potential corresponding to Eq.~(\ref{eq:ex_elegant}) is given by
\begin{eqnarray}\label{eq:vxhom}
v^{\rm hom}_{\rm x}(n) &\equiv& 
\frac{\partial [n\delta \varepsilon_{\rm x}(n)]}{\partial n}=\frac{3}{2} \varepsilon_{\rm F} \alpha_{\rm gr} F(\Lambda)\nonumber\\
&+&\varepsilon_{\rm F} \alpha_{\rm gr} \frac{\partial F}{\partial \Lambda}\times n\frac{\partial \Lambda}{\partial n}
\end{eqnarray}
where 
\begin{equation}
n\frac{\partial \Lambda}{\partial n}=
-\frac{1}{2}\Lambda~.
\end{equation}
We have chosen the following simple formula for $F(\Lambda)$ to parametrize the data in Ref.~\onlinecite{barlas_prl_2007}: 
\begin{equation}\label{eq:choice}
F(\Lambda)=\frac{1}{6g}\ln(\Lambda)+\frac{a_{\rm e}}{1+b_{\rm e}~\Lambda^{c_{\rm e}}}~,
\end{equation}
where the first term, which is the leading contribution in the limit $\Lambda \gg 1$, has been calculated analytically in Ref.~\onlinecite{barlas_prl_2007}. This term is largely responsible for the quasiparticle velocity enhancement in doped graphene sheets~\cite{barlas_prl_2007,ourdgastheory}. The numerical constants $a_{\rm e}, b_{\rm e}$, and $c_{\rm e}$
are given by
\begin{equation}
\label{eq:parameters}
\left\{
\begin{array}{l}
a_{\rm e}=0.0173671\\
b_{\rm e}=3.6642\times 10^{-7}\\
c_{\rm e}=1.6784
\end{array}
\right.~.
\end{equation}
Eq.~(\ref{eq:choice}) implies that
\begin{equation}
\frac{\partial F}{\partial \Lambda}=-\frac{a_{\rm e}b_{\rm e}c_{\rm e}}{(1+b_{\rm e}~\Lambda^{c_{\rm e}})^2}\frac{\Lambda^{c_{\rm e}}}{\Lambda}+\frac{1}{6g}\frac{1}{\Lambda}~.
\end{equation}
Note that for $n\to 0$ 
the exchange potential goes to zero like
\begin{eqnarray}
v^{\rm hom}_{\rm x}(n\to 0)\propto -{\rm sgn}(n)\alpha_{\rm gr}\sqrt{|n|}~\ln{|n|}~,
\end{eqnarray}
{\it i.e.} with an infinite slope.

\subsection{RPA correlation potential}
\label{sect:c}

The RPA correlation energy data of Ref.~\onlinecite{barlas_prl_2007} can be conveniently parametrized 
by the following formula
\begin{eqnarray}\label{eq:parametrization_correlation}
\frac{\delta \varepsilon^{\rm RPA}_{\rm c}(n)}{\varepsilon_{\rm F}}&=&
-\frac{\alpha^2_{\rm gr}}{6g}\xi(\alpha_{\rm gr})\ln{(\Lambda)}\nonumber\\
&+&
\frac{\alpha^2_{\rm gr}~a_{\rm c}(\alpha_{\rm gr})}{1+b_{\rm c}(\alpha_{\rm gr})\Lambda^{c_{\rm c}(\alpha_{\rm gr})}}
\end{eqnarray}
where
\begin{equation}
\left\{
\begin{array}{l}
a_{\rm c}(\alpha_{\rm gr})=-1/(63.0963+57.351226~\alpha_{\rm gr})\vspace{0.1cm}\\
b_{\rm c}(\alpha_{\rm gr})=(7.75095-0.08371~\alpha_{\rm gr}^{1.61167})\times 10^{-7}\vspace{0.1cm}\\      
{\displaystyle c_{\rm c}(\alpha_{\rm gr})=1.527+0.0239~\alpha_{\rm gr}-0.001201~\alpha_{\rm gr}^2}
\end{array}
\right.
\end{equation}
and
\begin{eqnarray}
\xi(\alpha_{\rm gr})&=&\frac{1}{2}\int_0^{+\infty}
\frac{dx}{(1+x^2)^{2}(\sqrt{1+x^2}+\pi\alpha_{\rm gr}/8)}.
\end{eqnarray}
Once again, the logarithmic contribution in Eq.~(\ref{eq:parametrization_correlation}) represents the leading term in the limit $\Lambda \gg 1$ and was calculated analytically in Ref.~\onlinecite{barlas_prl_2007}.

\begin{figure}
\begin{center}
\includegraphics[width=1.00\linewidth]{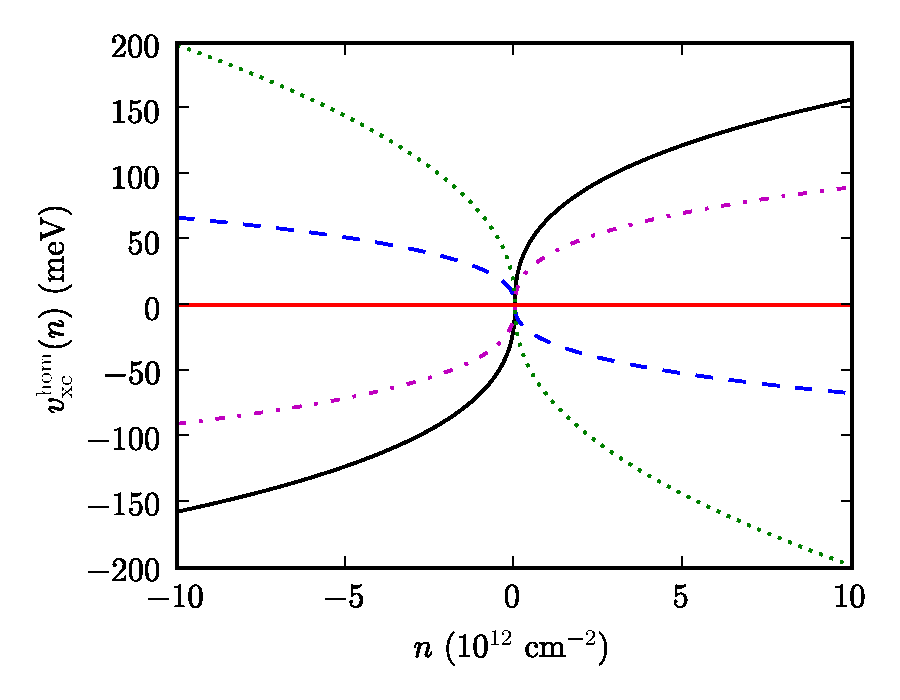}
\includegraphics[width=1.00\linewidth]{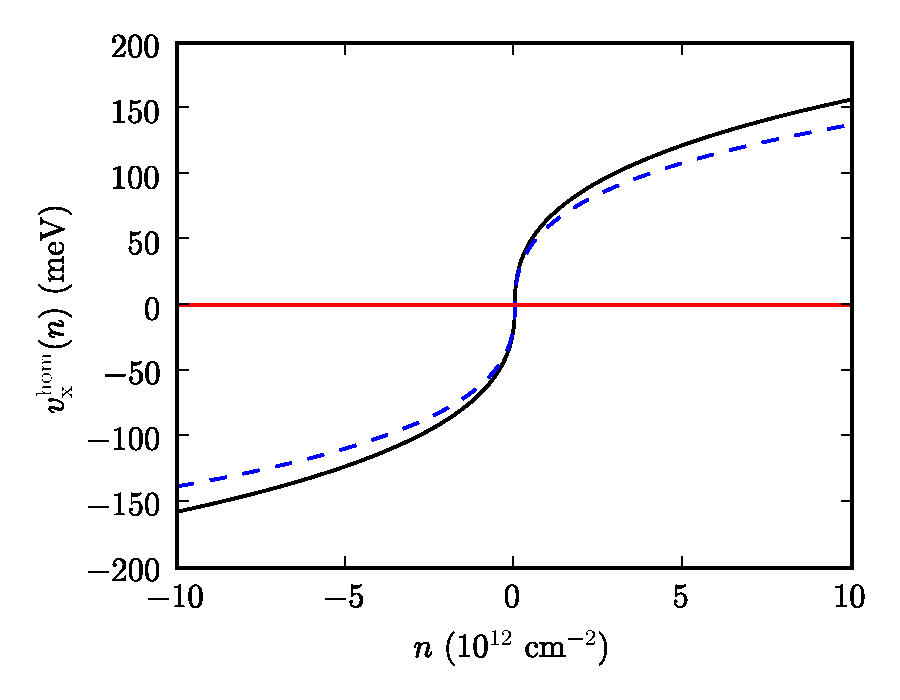}
\caption{(Color online) Top panel: 
the exchange and  RPA correlation potentials, $v^{\rm hom}_{\rm x}(n)$ [(black) solid line)] and $v^{\rm hom}_{\rm c}(n)$ [(blue) dashed line)], (in ${\rm meV}$ units) as functions of the density $n$ (in units of $10^{12}~{\rm cm}^{-2}$) for $\alpha_{\rm ee}=0.5$. 
Note how for $n\to 0$ both potentials have an infinite slope. 
The (magenta) dash-dotted line represents the full exchange-correlation 
potential, $v^{\rm hom}_{\rm xc}(n)=v^{\rm hom}_{\rm x}(n)+v^{\rm hom}_{\rm c}(n)$. The (green) dotted line is the quantum Monte Carlo exchange-correlation potential of a standard parabolic-band 2D electron gas~\cite{attaccalite_prl_2002}. For convenience we have chosen parameters corresponding to a 2D electron gas on a background with dielectric costant 
$4$ and with band mass $0.067~m$, $m$ being the electron mass in vacuum.
Bottom panel: the full exchange potential [(black) solid line)] is compared with its $\ln$-only approximation, 
[(blue) dashed line], {\em i.e.} retaining only the first term in Eq.~(\ref{eq:choice}).\label{fig:one}}
\end{center}
\end{figure}

Note that we can write Eq.~(\ref{eq:parametrization_correlation}) in the form:
\begin{equation}\label{eq:ec_elegant}
\delta \varepsilon^{\rm RPA}_{\rm c}=\varepsilon_{\rm F} \alpha^2_{\rm gr} G_{\alpha_{\rm gr}}(\Lambda)
\end{equation}
with
\begin{equation}
G_{\alpha_{\rm gr}}(\Lambda)=-\frac{\xi(\alpha_{\rm gr})}{6g}\ln(\Lambda)+\frac{a_{\rm c}(\alpha_{\rm gr})}{1+b_{\rm e}(\alpha_{\rm gr})~\Lambda^{c_{\rm e}(\alpha_{\rm gr})}}~.
\end{equation}
Following the same procedure highlighted above for the exchange contribution, one easily finds the correlation contribution to 
$v^{\rm hom}_{\rm xc}(n)$. The only necessary input to calculate this contribution is 
\begin{equation}
\frac{\partial G_{\alpha_{\rm gr}}}{\partial \Lambda}=-\frac{a_{\rm c}b_{\rm c}c_{\rm c}}{(1+b_{\rm c}~\Lambda^{c_{\rm c}})^2}\frac{\Lambda^{c_{\rm c}}}{\Lambda}-\frac{\xi(\alpha_{\rm gr})}{6g}\frac{1}{\Lambda}~.
\end{equation}
In the limit $n\to 0$ we find 
\begin{eqnarray}
v^{\rm hom}_{\rm c}(n\to 0) \propto {\rm sgn}(n)\alpha^2_{\rm gr}\xi(\alpha_{\rm gr})\sqrt{|n|}~\ln{|n|}~.
\end{eqnarray}
A plot of the exchange-correlation potential as a function of the density $n$ is given in Fig.~\ref{fig:one}. 
For the sake of comparison, in Fig.~\ref{fig:one} we also have plotted the quantum Monte Carlo 
exchange-correlation potential of the parabolic-band 2D electron gas~\cite{attaccalite_prl_2002}, 
after having antisymmetrized it for $n<0$.  We can clearly see from this plot how the density-dependence of the 
exchange-correlation potential of a uniform 2D liquid of massless Dirac fermions is precisely opposite 
to the one familiar from the ordinary LDA for parabolic-band inhomogeneous electron liquids. 
While the latter is negative for positive density, favoring inhomogeneous densities,
the former increases the energy cost of density increases, favoring more homogeneous densities and 
enhancing screening.  It is also apparent from this figure that the density dependence of the exchange-correlation potential 
can in some circumstances lead to effects which can give the appearance of a gap in the graphene sheet's Dirac bands. 
 
\section{Kohn-Sham-Dirac equation solutions: plane-wave method}
\label{sect:pw}

In this Section we discuss Kohn-Sham-Dirac equation solutions 
based on a supercell method and plane-wave expansions.
We consider massless Dirac fermions in a 2D (square) box of size $L\times L$ with {\it periodic} boundary conditions. 
In this case the Kohn-Sham-Dirac equations (\ref{eq:ksd}) can be conveniently 
solved by expanding the spinors $\Phi_\lambda({\bm r})$ in a plane-wave basis.
We discretize real space: ${\bm r} \to {\bm r}_{ij}=(x_{i},y_{j})$, $x_{i}=i\delta x$, $y_{j}=j\delta y$ with $i=1...N_x$ and $j=1...N_y$. Here $\delta x \times  N_{x}=\delta y \times N_{y}=L$. Fourier transforms ${\widetilde f}({\bm k})$ of real-space functions $f({\bm r})$ are calculated by means of a standard fast-Fourier-transform algorithm~\cite{website}
that allows us to compute ${\widetilde f}$ on the set of discrete wavevectors ${\bm k}_{ij}$,
\begin{equation}
{\bm k}_{ij}= (k_{x,i},k_{y,j}) 
= \frac{2\pi}{L}~(n_{x,i}, n_{y,j})~,
\end{equation}
with $-N_{x}/2 \le n_{x,i} < N_{x}/2 $ and $-N_{y}/2 \le n_{y,j} < N_{y}/2 $ (or, equivalently, $0 \le n_{x,i} < N_{x}$ and $0 \le n_{y,j} < N_{y}$), that belong to the Bravais lattice of the discretized box.
The definition of the Fourier transform that we use is the following:
\begin{equation}
\left\{
\begin{array}{l}
{\displaystyle f({\bm r})  =  \int \frac{d^{2}{\bm k}}{(2\pi)^{2}}~{\widetilde f}({\bm k})~e^{i{\bm k}\cdot{\bm r}}}
\vspace{0.1 cm}\\
{\displaystyle {\widetilde f}({\bm k})  =  \int d{^{2}\bm r}~f({\bm r})~e^{-i{\bm k}\cdot{\bm r}}}
\end{array}
\right.~.
\end{equation}
After discretization $f({\bm r})\to f_{ij}=f({\bm r}_{ij})$, 
${\widetilde f}_{ij}=\tilde{f}({\bm k}_{ij})$ with
\begin{eqnarray}
f_{ij}  &=&  \frac{1}{L^{2}}
\sum_{n=0}^{N_{x}-1}\sum_{m=0}^{N_{y}-1}~{\widetilde f}_{nm}~e^{i {\bm k}_{nm}\cdot{\bm r}_{ij}}
\end{eqnarray}
and
\begin{eqnarray}
{\widetilde f}_{ij} & = & L^{2} \times \frac{1}{N_{x}N_{y}}
\sum_{n=0}^{N_{x}-1}\sum_{m=0}^{N_{y}-1}~f_{nm}~e^{-i {\bm k}_{ij}\cdot{\bm r}_{nm}}~.
\end{eqnarray}
In all the numerical calculations reported on below we use $L$ as unit of length, 
$2\pi \hbar/L$ as the unit of momentum, and $\hbar v/L$ as the unit of energy.
In what follows we also set $\hbar=1$.

In momentum space Eq.~(\ref{eq:ksd}) reads
\begin{equation}\label{eq:ksd_momentum}
\sum_{{\bm k}'}\langle {\bm k} | [v {\bm \sigma}\cdot {\bm p} +  {\mathbb I}_\sigma V_{\rm KS}({\bm r})] |{\bm k}'\rangle {\widetilde \Phi}_{\lambda}({\bm k}')
=\varepsilon_{\lambda} {\widetilde \Phi}_{\lambda}({\bm k})~.
\end{equation}
Here $\lambda$ labels the eigenvalues of the Kohn-Sham-Dirac matrix 
${\cal H}^{\rm KSD}_{{\bm k},{\bm k}'}\equiv \langle {\bm k} |[v {\bm \sigma}\cdot {\bm p} + {\mathbb I}_\sigma 
V_{\rm KS}({\bm r})]|{\bm k}'\rangle$. 
The matrix elements of the kinetic Hamiltonian are given by
\begin{eqnarray}
\langle {\bm k}|~v {\bm \sigma} \cdot {\bm p}~|{\bm k}'\rangle &=&
v{\bm \sigma} \cdot {\bm k}' \delta_{{\bm k}, {\bm k}'}~.
\end{eqnarray}
We employ a momentum space cut-off $k_{x,i},k_{y,j} \in [-k_{\rm c},+k_{\rm c}]$ which 
does not exceed the Brillouin-zone boundary defined by our real-space discretization:  
$k_{\rm c} < \pi / \delta x, \pi/\delta y$.  $k_{\rm c}$ defines the range of momenta used in the expansion 
of the Hamiltonian ${\cal H}^{\rm KSD}_{{\bm k},{\bm k}'}$ and thus defines its dimension $d_{\rm H}$:
\begin{equation}
d_{\rm H} = 2\times \left(2\times \frac{Lk_{\rm c}}{2\pi}+1\right)^{2}~.
\end{equation}
The factor of $2$ here is due to the sublattice pseudospin degree of freedom.  As already stated above, 
real spin and valley degrees of freedom enter our calculations only through the trivial degeneracy factors they imply.
Given a value of $k_{\rm c}$ the Kohn-Sham-Dirac matrix ${\cal H}^{\rm KSD}_{{\bm k},{\bm k}'}$ 
has $d_{\rm H}$ eigenvalues, labeled by the discrete index $\lambda=1,\dots,d_{\rm H}$.

\section{Non-linear screening of Coulomb impurities}
\label{sect:nls}

As an illustration we apply the LDA-DFT method described above 
to study the non-linear screening of $N_{\rm imp} \geq 1$ point-like impurities with charge $Ze$ 
($Z$ can be either positive and negative and $e>0$ in this work) located at random positions on a plane at 
a distance $d$ from the 2D chiral electron gas (CEG) plane.  The approximately 
linear dependence of conductivity on carrier density in graphene sheets suggests\cite{austincoulomb,marylandcoulomb} that 
nearby charged impurities are the dominant disorder source in most current graphene samples.

\subsection{Constructing the KS potential and the ground-state density}
\label{eq:construction}

We assume that the 2D CEG has a spatially averaged 
$\pi$-electron density 
\begin{equation}\label{eq:average}
n_0=\frac{2}{{\cal A}_0}+ {\bar n}_{\rm c}~.
\end{equation}
Here $2/{\cal A}_0$ is the density of a neutral graphene sheet and ${\bar n}_{\rm c}$ is the spatially averaged carrier density,  
which can be positive or negative and controlled by gate voltages~\cite{reviews,martin_et_al,berkely_MM2008}.
In what follows we write ${\bar n}_{\rm c} \equiv 4{\cal Q}/L^2$, where ${\cal Q}$ is the
number of carriers per spin and valley in our supercell.  Because of the role played by gate voltages in 
experiment, there is no reason to impose a charge-neutrality relationship between the number of impurities 
$N_{\rm imp}$ and ${\cal Q}$. 

The external potential $V_{\rm ext}({\bm r})$
is given by:
\begin{equation}\label{eq:ext_pot}
V_{\rm ext}({\bm r})=-\sum_{i=1}^{N_{\rm imp}}\frac{Ze^2}{\epsilon \sqrt{|{\bm r}-{\bm R}_i|^2 + d^2}}~,
\end{equation}
where ${\bm R}_i$ are random positions in the supercell. 
For simplicity, all charges have been taken to have the same $Z$ in Eq.~(\ref{eq:ext_pot}). 
The matrix elements of the disorder potential in Eq.~(\ref{eq:ext_pot}) are given by
\begin{eqnarray}
\langle {\bm k}|V_{\rm ext}({\bm r})|{\bm k}'\rangle &=&
{\widetilde V}_{\rm ext}({\bm k}-{\bm k}')~{\cal F}_{\rm imp}({\bm k}-{\bm k}')~,
\end{eqnarray}
where ${\widetilde V}_{\rm ext}({\bm q})=-2\pi Z e^2 \exp{(-q d)}/(\epsilon q)$ 
is the Fourier transform of the potential created by a single impurity and
\begin{equation}
{\cal F}_{\rm imp}({\bm k}-{\bm k}')=\frac{1}{L^2}
\sum_{i=1}^{N_{\rm imp}}e^{-i({\bm k}-{\bm k}')\cdot {\bm R}_i}
\end{equation}
is a geometric form factor that depends only on the positions of the impurities. 
The impurity charges are replicated in each supercell and the total potential $V_{\rm ext}({\bm r})$
therefore has the supercell periodicity. 
We set ${\widetilde V}_{\rm ext}({\bm k}={\bm k}')=0$, thereby choosing the 
zero of energy at the Dirac-point energy in the spatially averaged external potential.

The ground-state density profile $n({\bm r})$ in the external potential given by Eq.~(\ref{eq:ext_pot}) is computed from Eq.~(\ref{eq:density}) by summing over $\lambda=1,\dots,\lambda_{\rm max}$, where 
the KS energy levels are arranged in ascending order, 
$\varepsilon_{1} \leq \dots \leq \varepsilon_{\lambda_{\rm max}} \leq \dots \leq \varepsilon_{d_{\rm H}}$. 
Since half of the system's $\pi$-orbitals are occupied in a 
neutral graphene sheet, $\lambda_{\rm max}$ is related to the
average $\pi$-electron density of the graphene sheet $n_0=4 (d_{\rm H}/2 + {\cal Q})/L^2$ by  
\begin{eqnarray}\label{eq:final_lambdamax}
\lambda_{\rm max}=  \frac{d_{\rm H}}{2}+{\cal Q}~.
\end{eqnarray}
Note that this implies the following relationship between the momentum-space cutoff $k_{\rm c}$ 
and the area of the system $L^2$ in units of ${\cal A}_0$: $L^2/{\cal A}_0=2~[2 Lk_{\rm c}/(2\pi)+1]^2$.
In our self-consistent numerical calculations we evaluate only the 
deviation of the density from its average value in the supercell: 
\begin{eqnarray}\label{eq:renormalization}
\delta n({\bm r})=n({\bm r})-n_0~.
\end{eqnarray}
The corresponding quantity in momentum space $\delta {\widetilde n}({\bm k})$  
is given by $\delta {\widetilde n}({\bm k})={\widetilde n}({\bm k})-n_0\delta_{{\bm k}, {\bm 0}}$. 
Note that $\delta n({\bm r})$ is {\it charge~neutral}, {\it i.e.} $\delta {\widetilde n}({\bm k}={\bm 0})=0$.
The matrix elements of the Hartree term in the Kohn-Sham-Dirac equation are given by
\begin{eqnarray}\label{eq:hartree_momentumspace}
\langle{\bm k}| \Delta V_{\rm H}({\bm r}) |{\bm k}'\rangle=
\frac{2\pi e^{2}}{\epsilon |{\bm{k}-\bm{k}'}|}\; \delta {\widetilde n}({\bm k} - {\bm k}')~.
\end{eqnarray}

The matrix elements of the exchange-correlation potential can be calculated numerically from
\begin{equation}\label{eq:lda_momentum_space}
\langle{\bm k}| V_{\rm xc}({\bm r}) |{\bm k}'\rangle =\frac{1}{L^2}
\int d^2 {\bm r}~V_{\rm xc}({\bm r})~e^{-i({\bm k}-{\bm k}')\cdot {\bm r}}~,
\end{equation}
where $V_{\rm xc}({\bm r})$ is given by Eq.~(\ref{eq:lda}) with the carrier density
\begin{equation}
n_{\rm c}({\bm r})=n({\bm r})-
\frac{2 d_{\rm H}}{L^2} =  \delta n({\bm r}) + \frac{4{\cal Q}}{L^2}~.
\end{equation}

\subsection{Numerical results}

In this Section we report some illustrative numerical results that we have obtained 
applying the LDA-DFT method described above. All the numerical results presented in this work 
were obtained with $\eta=1$ [see Eq.~(\ref{eq:eta}) for the definition of $\eta$].

\begin{figure}
\begin{center}
\tabcolsep=0cm
\begin{tabular}{cc}
\includegraphics[width=0.50\linewidth]{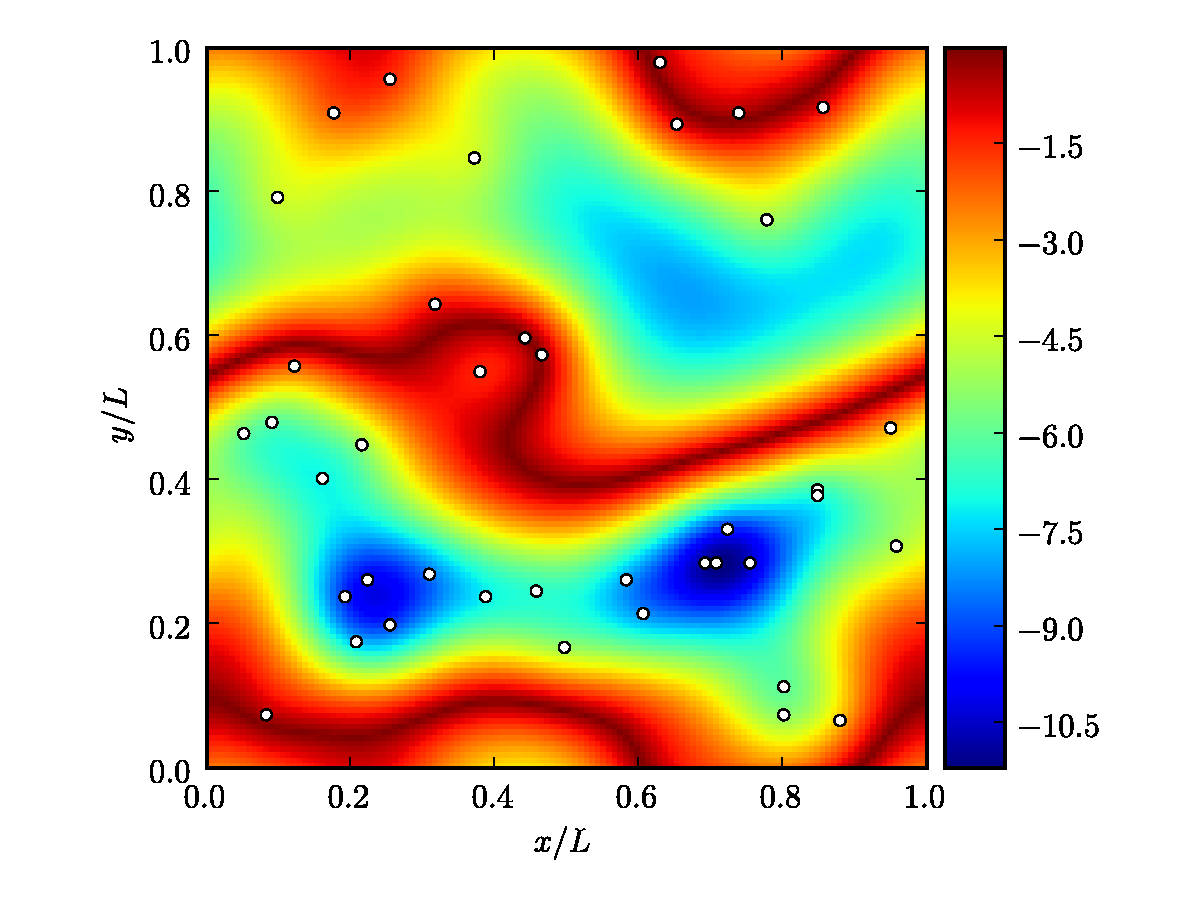}&
\includegraphics[width=0.50\linewidth]{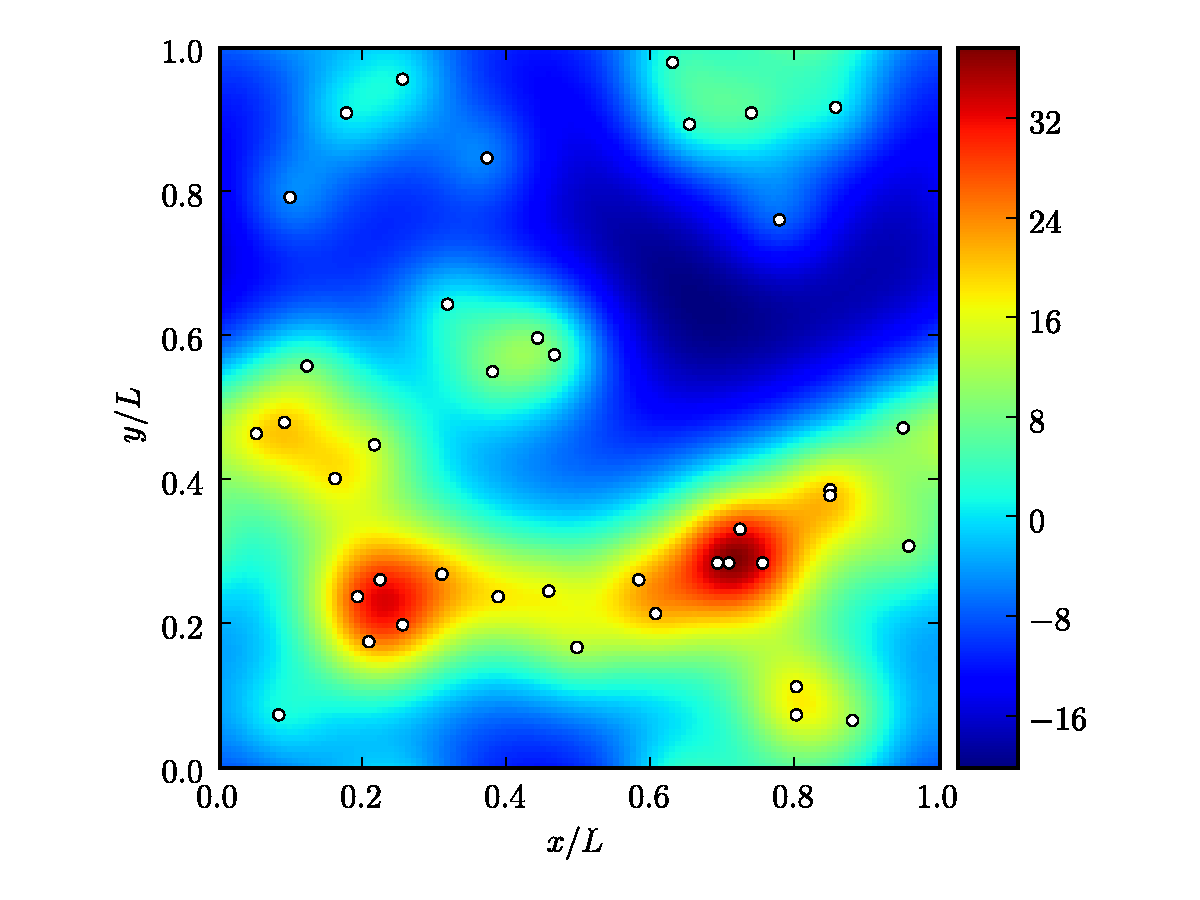}\\
\includegraphics[width=0.50\linewidth]{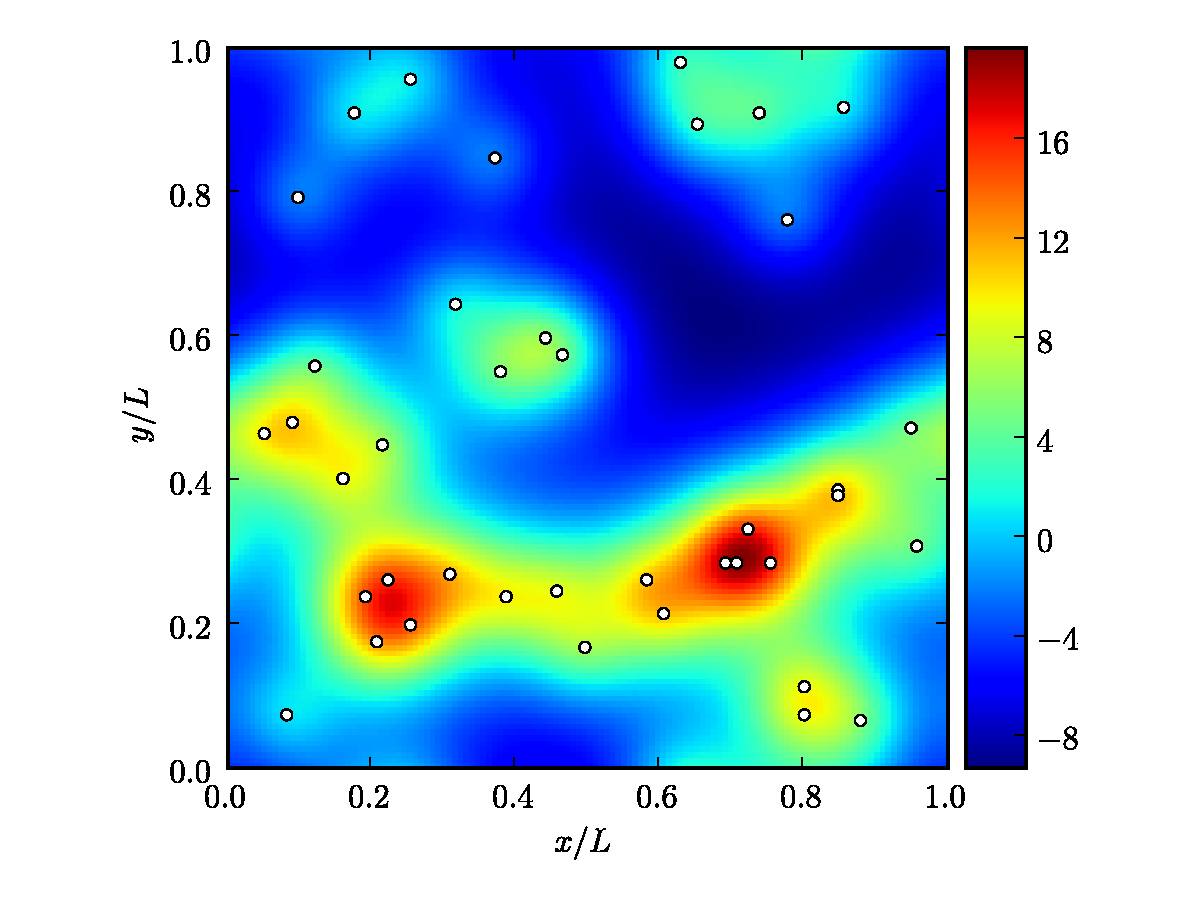}&
\includegraphics[width=0.50\linewidth]{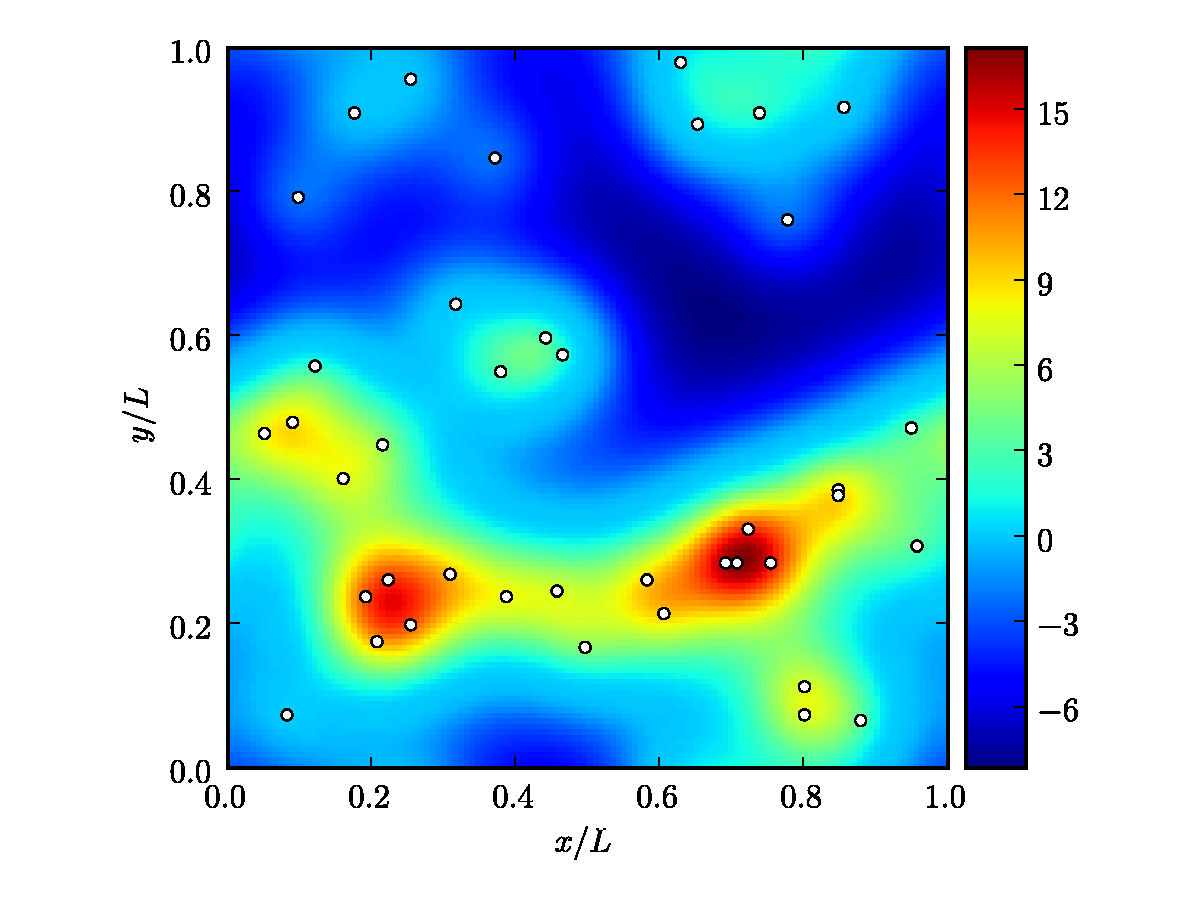}
\end{tabular}
\end{center}
\caption{(Color online)  Top left panel: a color plot of the external potential $V_{\rm ext}({\bm r})$ (in units of $\hbar v/L$) 
as a function of $x/L$ and $y/L$. The system parameters are $N_x=N_y=128$, $k_{\rm c}=(2\pi/L) \times 10$, $N_{\rm imp}=40$, $Z=+1$, $\alpha_{\rm ee}=0.5$, ${\cal Q}=0$, and $d/L=0.1$. The small circles represent the positions of the impurities for a particular realization of disorder. Top right panel: a color plot of the corresponding  non-interacting 
ground-state density profile $\delta n ({\bm r})$ (in units of $1/L^2$) as a function of $x/L$ and $y/L$. 
Bottom left panel: Hartree-only ground-state density profile. Bottom right panel: same as in the bottom left panel but with the addition of the exchange and RPA correlation potential.\label{fig:two}}
\end{figure}

In Fig.~\ref{fig:two} we illustrate the real-space density profile 
$\delta n({\bm r})$ of a neutral-on-average (${\cal Q}=0$) 2D CEG subjected to 
the external potential of $N_{\rm imp}=40$ impurities with $Z=+1$ 
located at a distance $d=0.1~L$ from the graphene plane 
[the corresponding external potential $V_{\rm ext}({\bm r})$ is illustrated in the top left panel of Fig.~\ref{fig:two}]. 
In this particular simulation we have used $\alpha_{\rm ee}=0.5$ and $k_{\rm c}=(2\pi/L)\times 10$, 
which corresponds to an effective square size $L^2=882~{\cal A}_0\sim 46~{\rm nm}^2$. 
The charges are therefore separated from the graphene layer 
by $d \sim 0.7~{\rm nm}$.  This model is motivated by growing experimental evidence that 
the dominant source of disorder in most graphene samples is external charges, probably 
located in the nearby substrate.

In Fig.~\ref{fig:two} we have reported: (i) the 
non-interacting Dirac electron density profile, which is obtained by setting the Hartree 
and exchange-correlation potentials in the Kohn-Sham-Dirac Hamiltonian to zero; (ii) the ``Hartree-only" density profile, which is obtained by solving the Kohn-Sham-Dirac equations self-consistently with $V_{\rm xc}({\bm r})=0$; and (iii) the ``full" density profile, which includes both Hartree and exchange-correlation effects. The self-consistent calculations are iterated until the Kohn-Sham potential is converged to a relative precision of $\sim 10^{-3}$. 

Electron-hole puddles, similar to those observed in Refs.~\onlinecite{martin_et_al,berkely_MM2008},
are evident in all these plots, although there are 
qualitative and quantitative differences between the non-interacting 
density profile and the ones that include electron-electron interactions 
(the experimental observation that the spatial pattern of electron-hole bubbles 
is not correlated with the topography of the graphene sheets~\cite{berkely_MM2008}, is consistent 
with the inference~\cite{austincoulomb,marylandcoulomb} from conductivity-{\em vs.}-carrier density data 
that remote charges rather than sheet corrugations dominate disorder). 
To begin with, we note how the inclusion of the Hartree term has the (expected) effect 
of reducing the amplitude of the spatial fluctuations of $\delta n({\bm r})$ quite dramatically, by 
approximately a factor of two in these non-linear screening calculations.  
It is interesting to compare this reduction factor with what would be expected in a linear screening approximation.
Neutral graphene has the unusual property that its static dielectric function $\varepsilon(q)$ 
neither diverges as wavevector $q$ goes to 
zero, as it would in a 2D metal, nor approaches $1$, as it would in a 2D semiconductor.  Instead 
\begin{equation} 
\varepsilon(q) = 1 - \frac{2\pi e^2}{\epsilon q}~{\widetilde \chi}_{\rho\rho}(q) 
\end{equation} 
approaches a constant because the polarization function 
${\widetilde \chi}_{\rho\rho}(q)$ (or {\it proper} density-density response 
function~\cite{Giuliani_and_Vignale}) has a non-analytic linear dependence on $q$
due to inter-band transitions with vanishing energy denominators.  
In the Hartree approximation 
[${\widetilde \chi}_{\rho\rho}(q) \to \chi^{(0)}(q)$, where $\chi^{(0)}(q)$ is the non-interacting polarization function~\cite{barlas_prl_2007}: 
see Sect.~5.3.1 of Ref.~\onlinecite{Giuliani_and_Vignale} for more details]
\begin{equation} 
\varepsilon(q) \to 1+ \frac{\pi}{8}~g\alpha_{\rm ee} \sim 1.78 
\end{equation} 
for the value of $\alpha_{\rm ee}$ used in our calculations.  
When exchange and correlations corrections are included in 
$\varepsilon(q)$ increases by a small fraction, enhancing screeening.
The influence of interactions on the non-linear screening calculations summarized in Fig.~\ref{fig:two} 
is therefore (perhaps surprisingly) broadly consistent with expectations based on linear screening 
theory - even at a semi-quantitative level.  
Qualitative non-linear effects do however appear in some details, as we now explain.

\begin{figure}
\begin{center}
\includegraphics[width=1.00\linewidth]{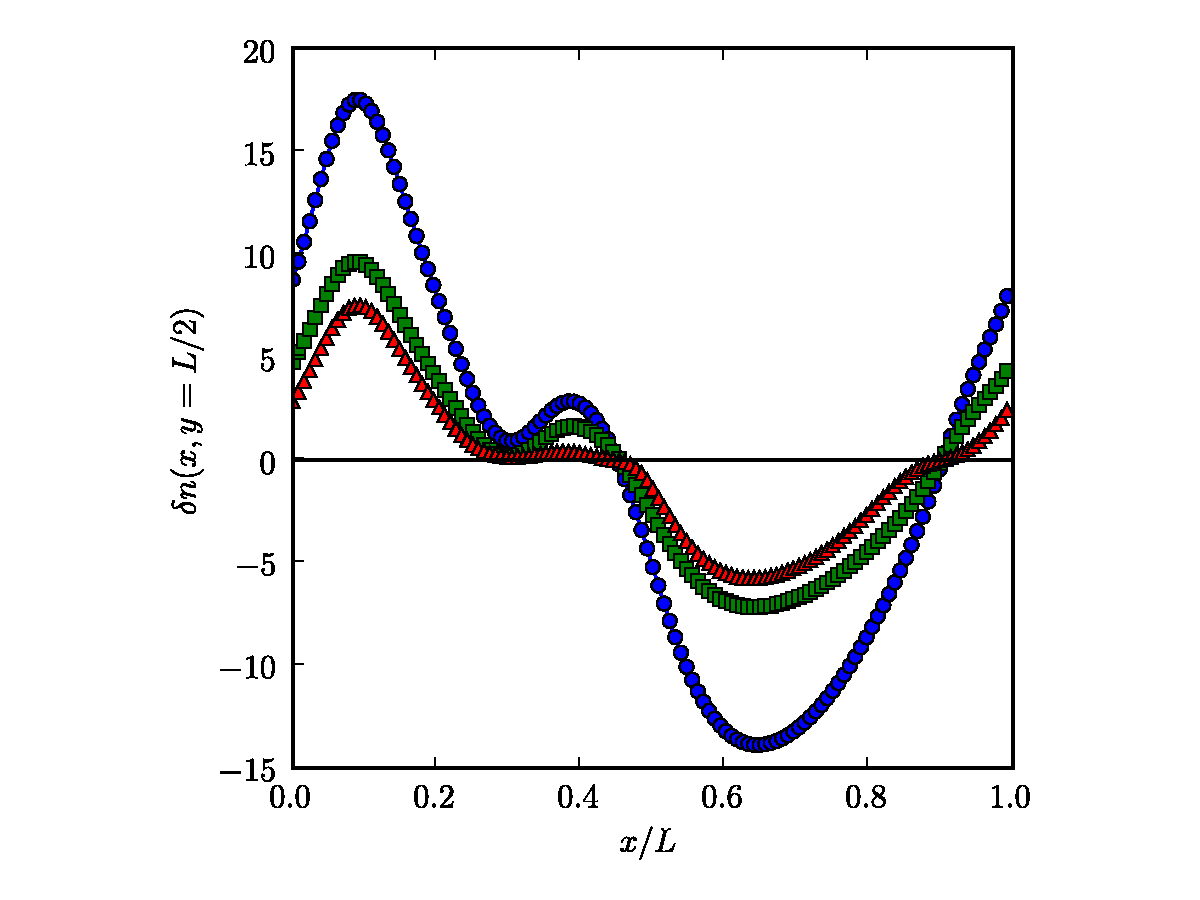}
\caption{(Color online). A one-dimensional plot of $\delta n ({\bm r})$ corresponding to the data in Fig.~\ref{fig:two} 
as a function of $x/L$ for $y/L=0.5$. The circles label the non-interacting result, the squares label the 
Hartree-only self-consistent result, and the triangles label the full self-consistent result.\label{fig:three}}
\end{center}
\end{figure}

In Fig.~\ref{fig:three} we examine the induced carrier density in more detail
by plotting $\delta n({\bm r})$ as a function of $x$ for a fixed value of $y$. 
Here we see clearly that $V_{\rm xc}({\bm r})$ tends to cause the 
density to vary less rapidly in those spatial regions at which the
carrier-density changes sign.  The origin of this behavior in our calculations 
is that the exchange-correlation potential increases especially rapidly with 
density in these regions.  This aspect of the induced density profile 
is similar to the behavior which would be produced by an energy-gap of 
$\sim 0.1~{\rm eV}$ in the graphene bands (see Fig.~\ref{fig:one}).  
The rapid change in exchange-correlation potential with density alters the 
statistical distribution of density-values in a disordered sample, as studied in some
detail using a Thomas-Fermi approximation for the non-interacting kinetic 
energy functional (and including local-density-approximation exchange)
by Rossi and Das Sarma in a recent paper~\cite{enrico_condmat}.
Thomas-Fermi theory is formally based on a gradient expansion of the total energy density (see {\it e.g.} Sect.~7.3.1 in 
Ref.~\onlinecite{Giuliani_and_Vignale}).  
When applied to graphene, assuming that the typical length scale for density variations in the 2D CES is the
inverse of the Thomas-Fermi screening wavevector $k_{\rm TF} = 2 \pi e^2 \nu(\varepsilon_{\rm F})/\epsilon$ 
[here $\nu(\varepsilon_{\rm F})=g \varepsilon_{\rm F}/(2\pi v^2)$ is the density-of-states at the Fermi energy],
Thomas-Fermi theory can be viewed as an expansion in powers of $k_{\rm TF}/k_{\rm F} = g \alpha_{\rm ee}$.  
As emphasized by Fogler and collaborators~\cite{fogler}, this parameter is not 
small when the value used for $\alpha_{\rm ee}$ is in the range $\sim 0.5$ thought to be appropriate for graphene on ${\rm SiO}_2$.  
In our approach we avoid a local-density-approximation for the 
non-interacting kinetic energy functional by solving microscopic 
Kohn-Sham-Dirac equations (this is the idea behind the Kohn-Sham mapping~\cite{dft}).  
We cannot avoid the local-density approximation 
for the exchange-correlation potential however [Eq.~(\ref{eq:lda})], and it must be acknowledged that this is a 
defect of our theory, and one that is not easily remedied.  The situation is similar to that 
in standard DFT applications, in which the local density-approximation is not rigorously valid 
on atomic length scales. It has nevertheless been possible to remedy defects of the 
local-density-approximation in many circumstances by using modified functionals, for example 
generalized-gradient approximations, which are often semi-phenomenological in character. 
Our expectation is that the LDA for exchange and correlation in graphene will improve 
accuracy compared to Thomas-Fermi approximation theories in which the band energy is also 
approximated using an LDA.
In addition, it will likely prove possible to compensate for the 
main-defects of the exchange-correlation
LDA by using modified exchange-correlation energy 
functionals which are informed by comparisons between theory and experiment.

\begin{figure}
\begin{center}
\tabcolsep=0cm
\begin{tabular}{cc}
\includegraphics[width=0.50\linewidth]{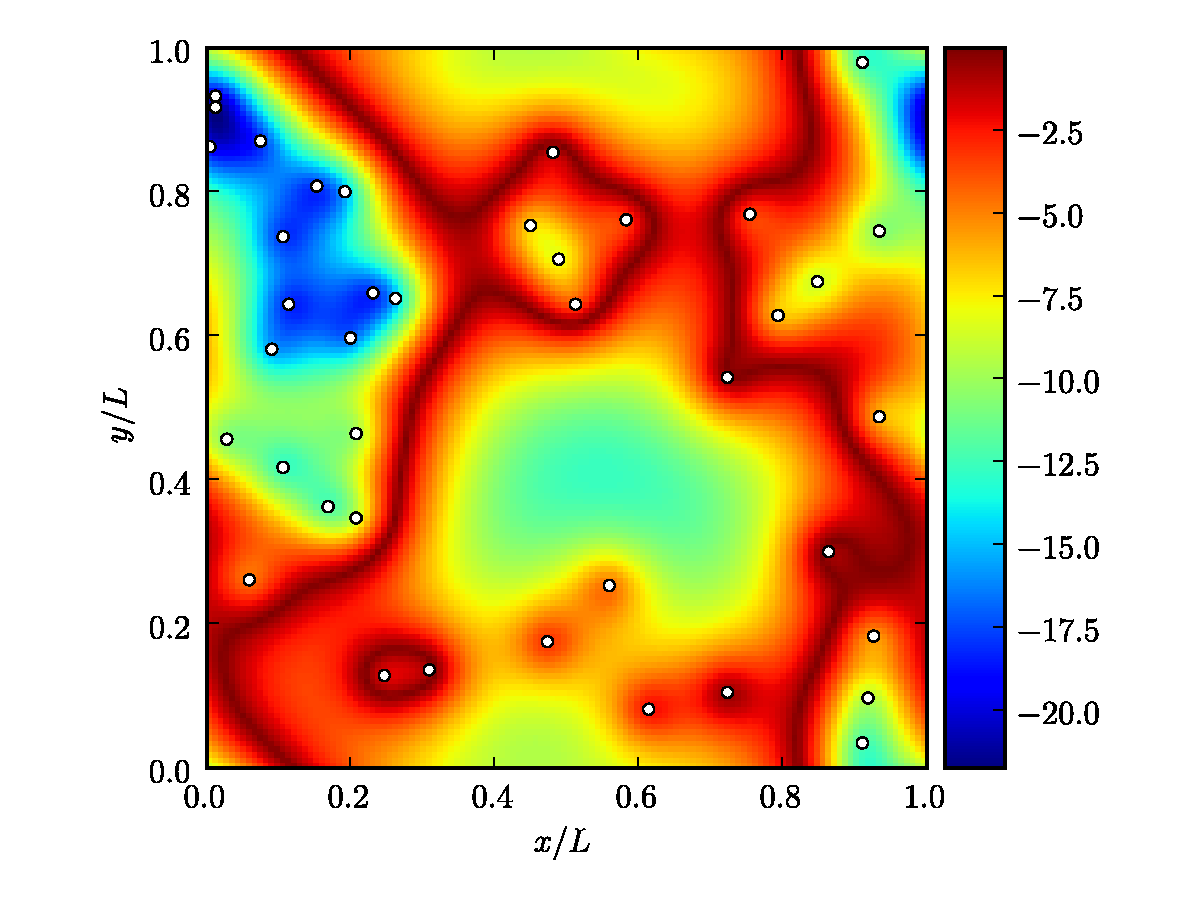}&
\includegraphics[width=0.50\linewidth]{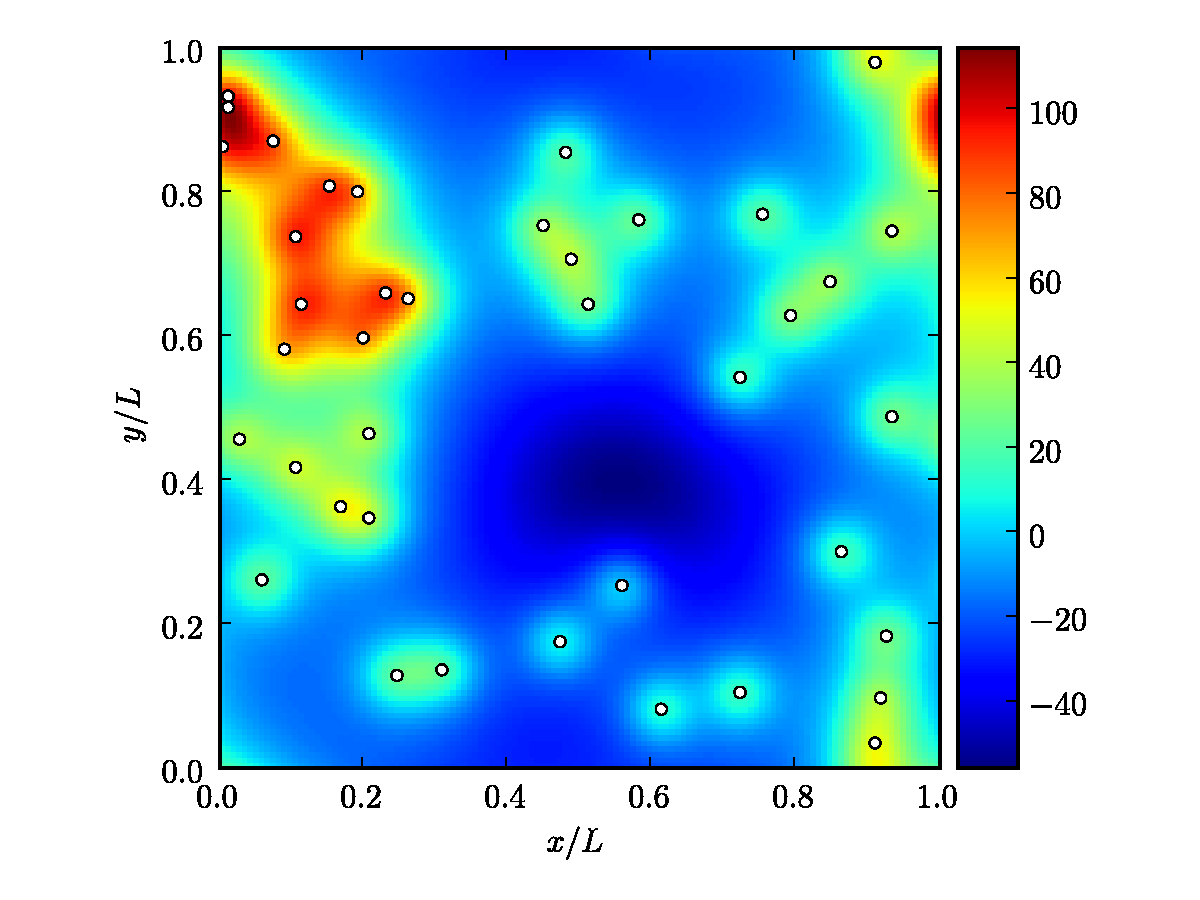}\\
\includegraphics[width=0.50\linewidth]{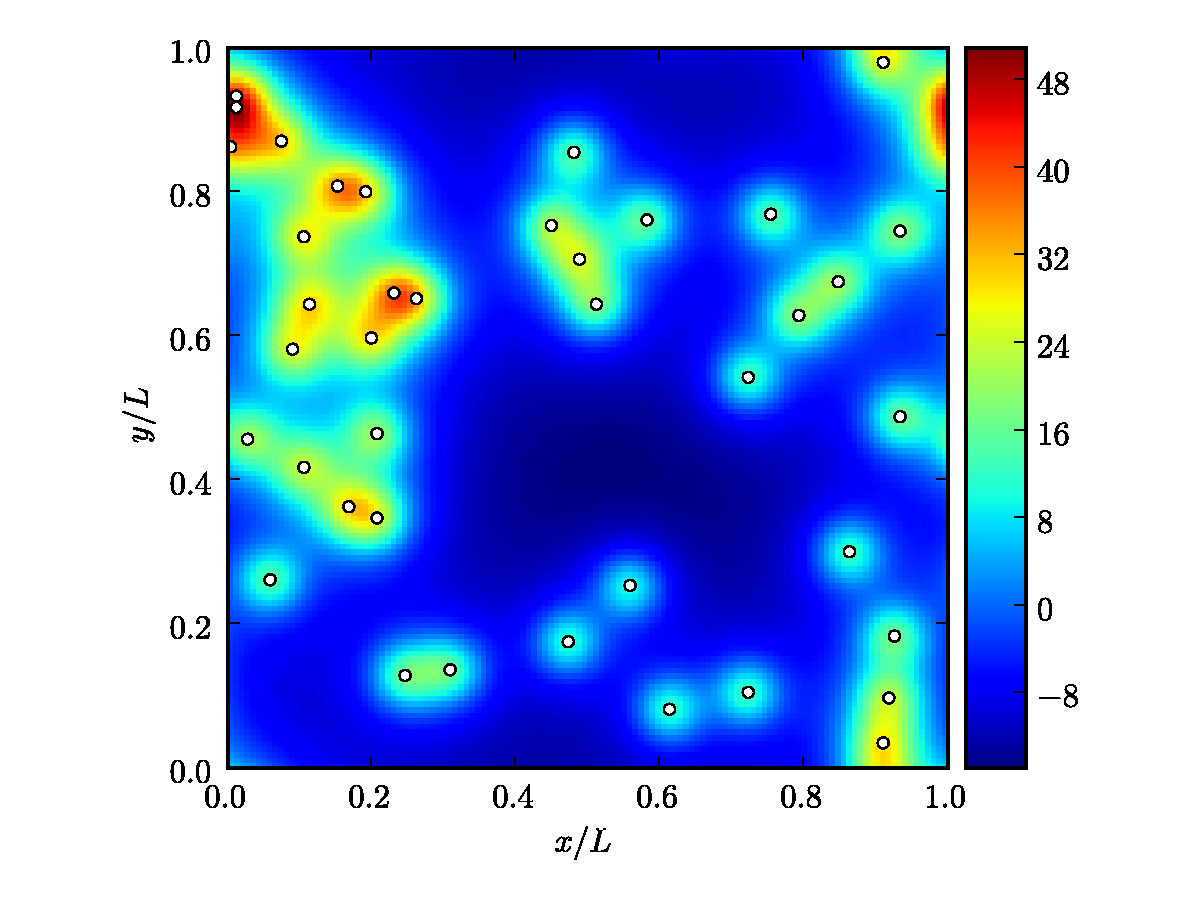}&
\includegraphics[width=0.50\linewidth]{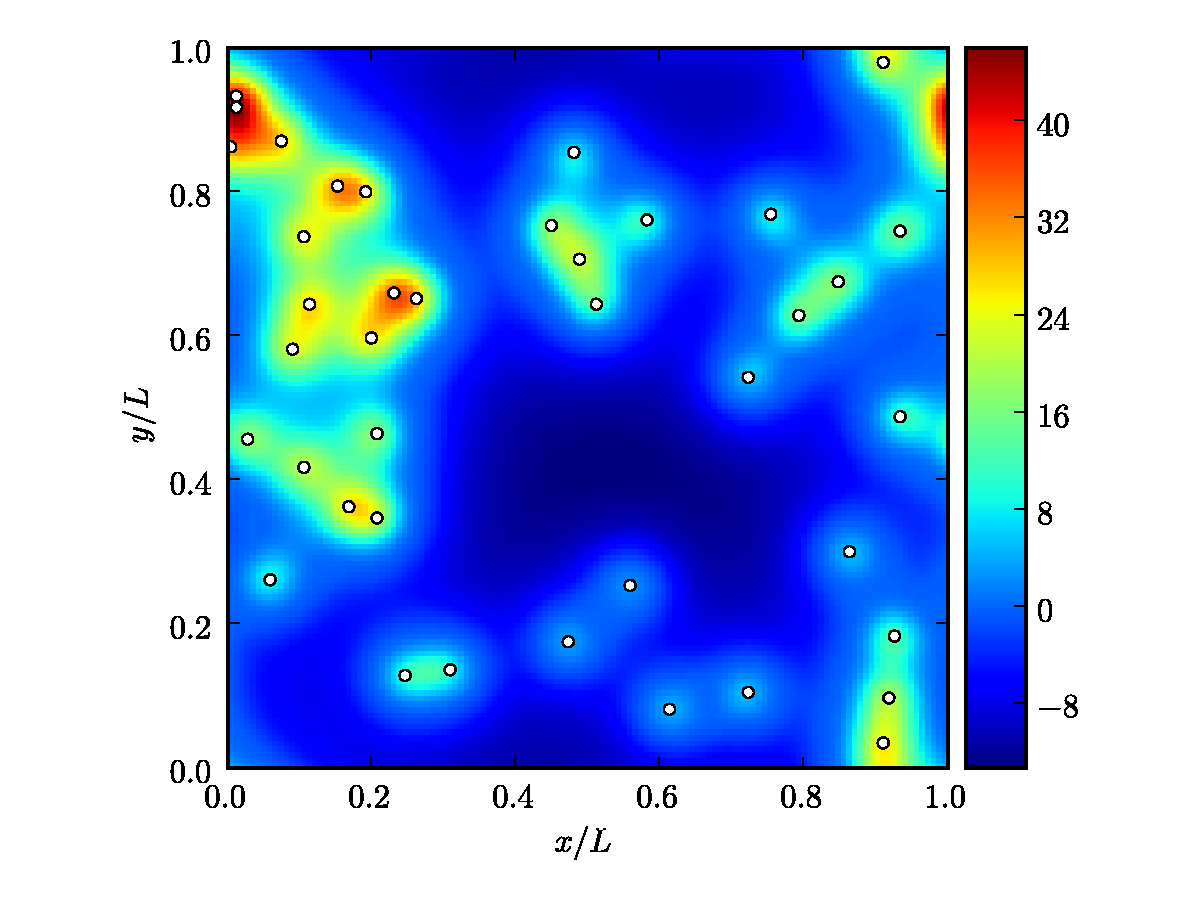}
\end{tabular}
\end{center}
\caption{(Color online) Same as in Fig.~\ref{fig:two} but for a different distribution of charges 
and for $d/L=0.05$ instead of $d/L=0.1$.\label{fig:four}}
\end{figure}

\begin{figure}
\begin{center}
\includegraphics[width=1.0\linewidth]{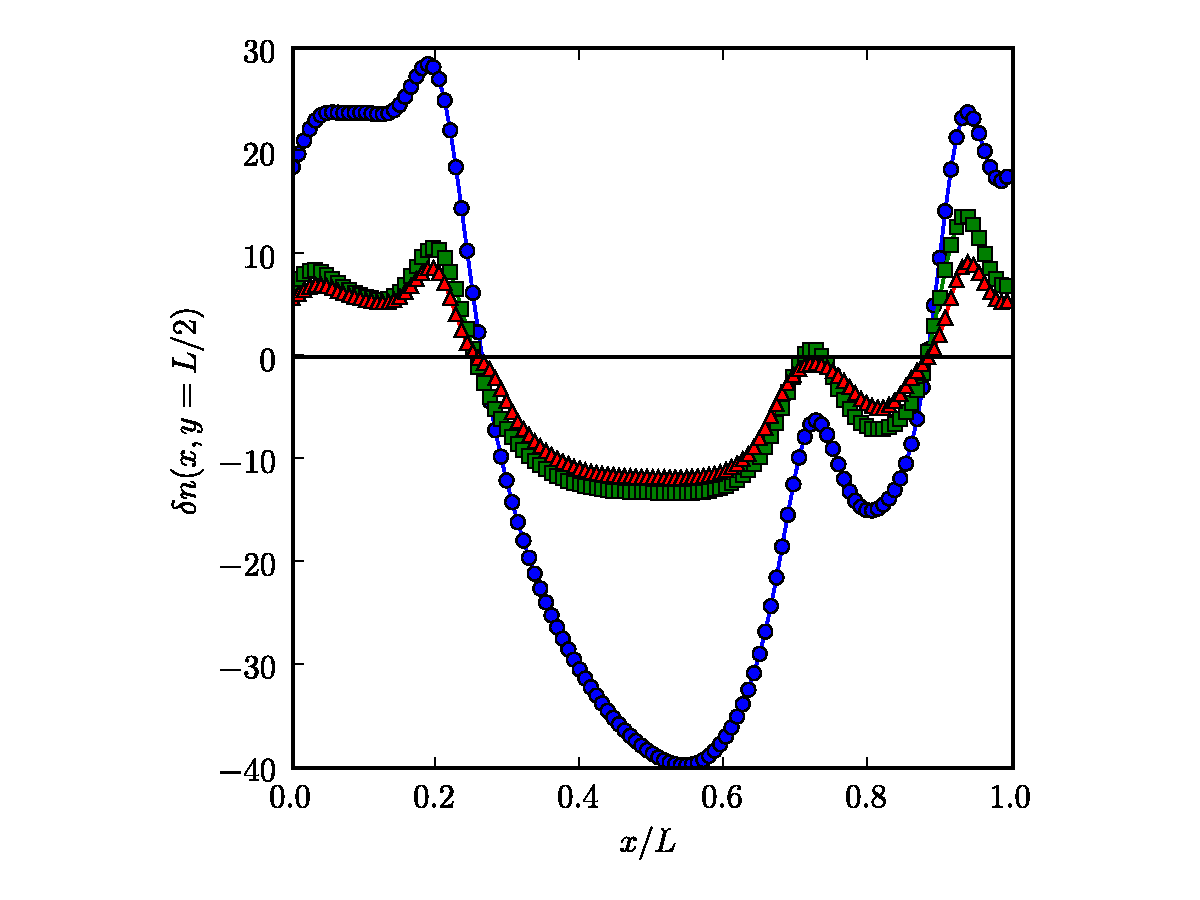}
\caption{(Color online). A one-dimensional plot of $\delta n ({\bm r})$ corresponding to the data in Fig.~\ref{fig:four} 
as a function of $x/L$ for $y/L=0.5$. The color coding is the same as in Fig.~\ref{fig:three}.\label{fig:five}}
\end{center}
\end{figure}

In Figs.~\ref{fig:four} and~\ref{fig:five} we report results similar to those presented 
in Figs.~\ref{fig:two} and~\ref{fig:three}, but for a separate realization of the random charged impurity distribution and a 
smaller separation between the impurity plane and the graphene plane, $d=0.05~L$.  When the impurities 
are closer to the graphene plane the role of the exchange-correlation potential 
seems to become less important.  
Conversely, for larger $d$ exchange and correlation effects increase in importance.
Because of the peculiar response of Dirac fermions, quite localized charge 
distributions can be induced by disorder potential features, even when those features are weak.
Indeed we find that for large separations between the graphene and impurity planes, the 
Kohn-Sham-Dirac equations do not always converge, indicating the possible importance in
some circumstances of correlation effects which cannot be captured by the KS LDA theory. 

\begin{figure}
\begin{center}
\tabcolsep=0cm
\begin{tabular}{cc}
\includegraphics[width=0.50\linewidth]{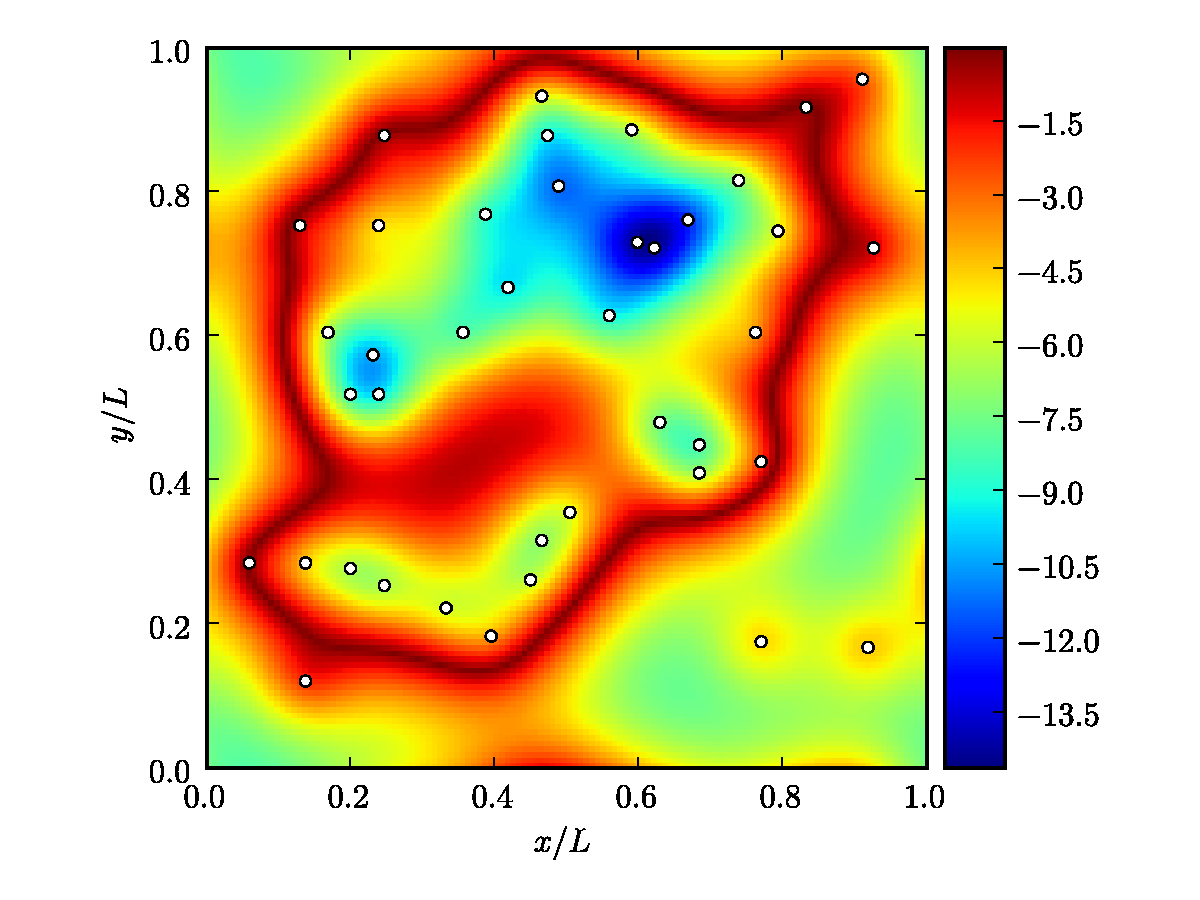}&
\includegraphics[width=0.50\linewidth]{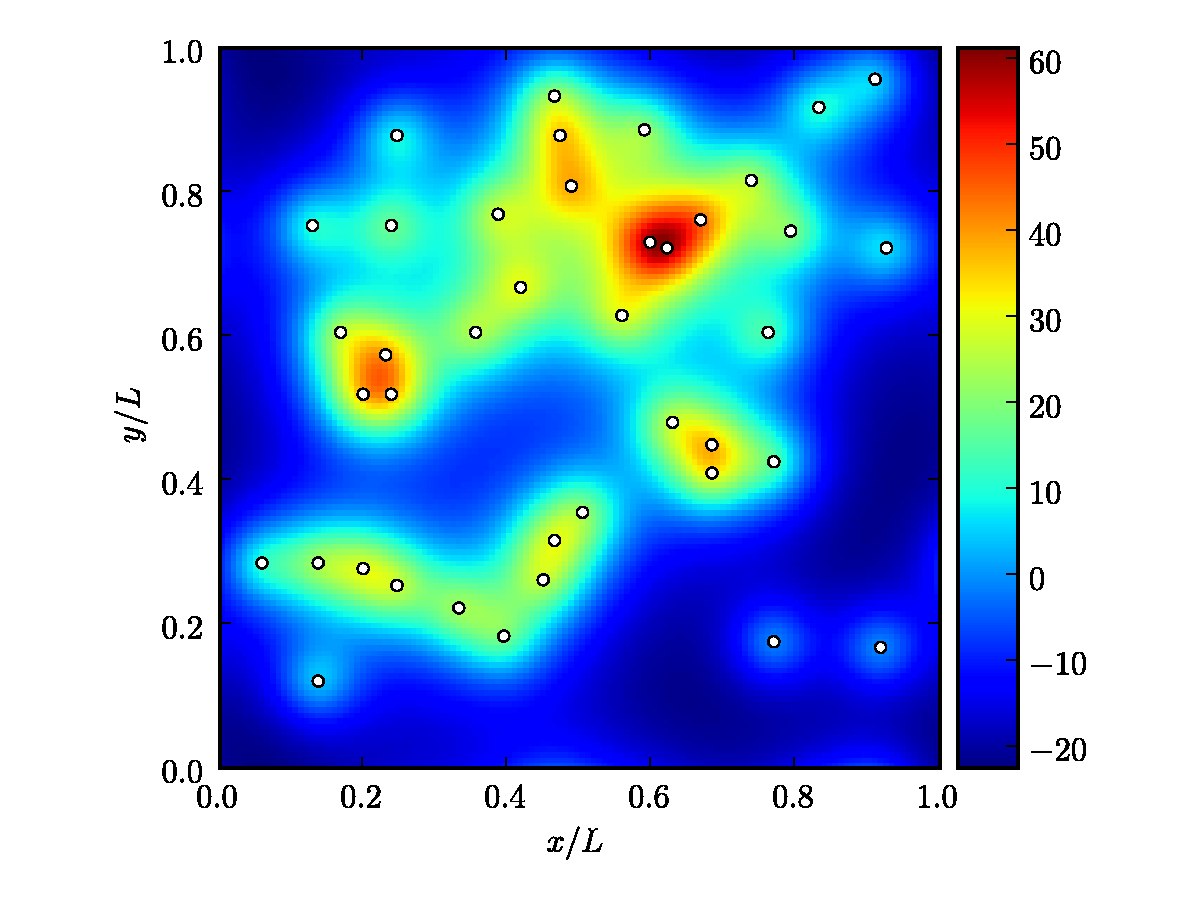}\\
\includegraphics[width=0.50\linewidth]{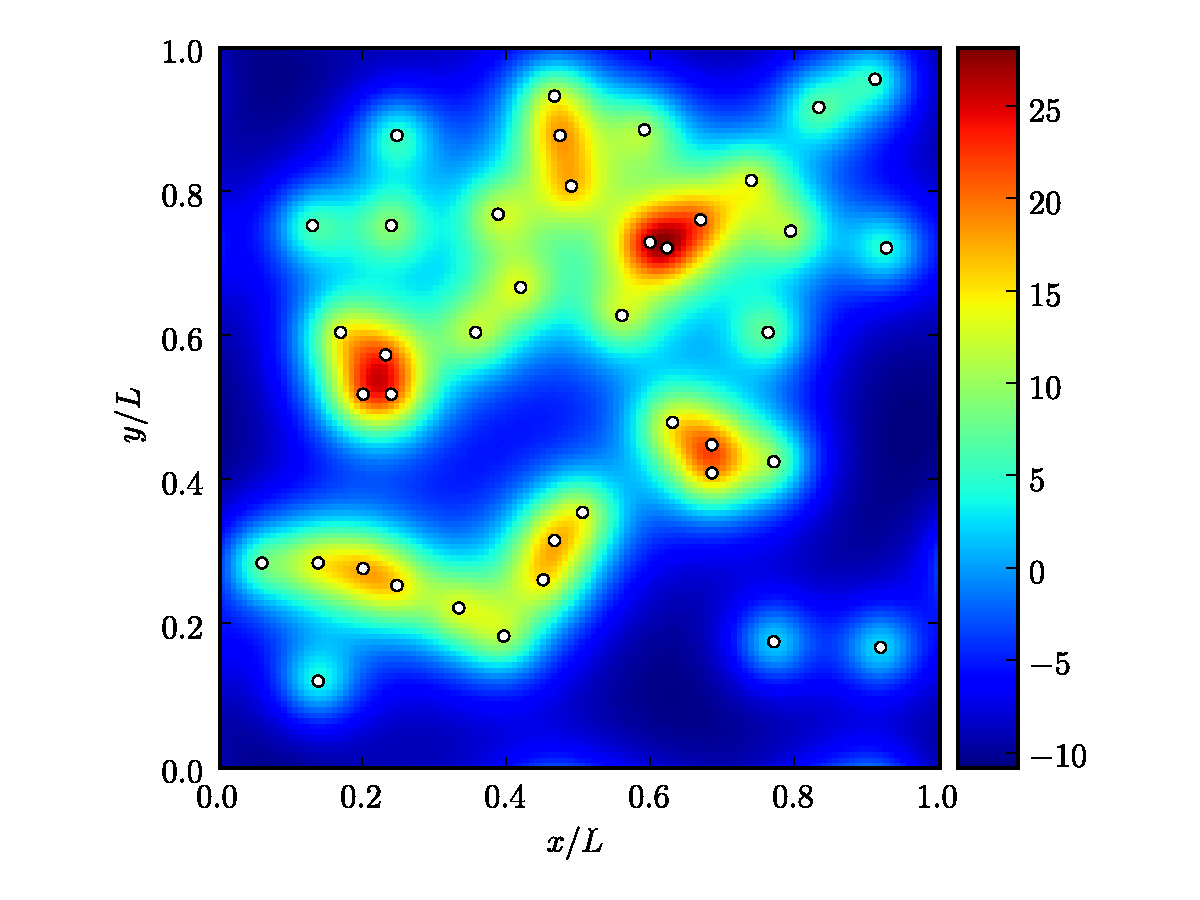}&
\includegraphics[width=0.50\linewidth]{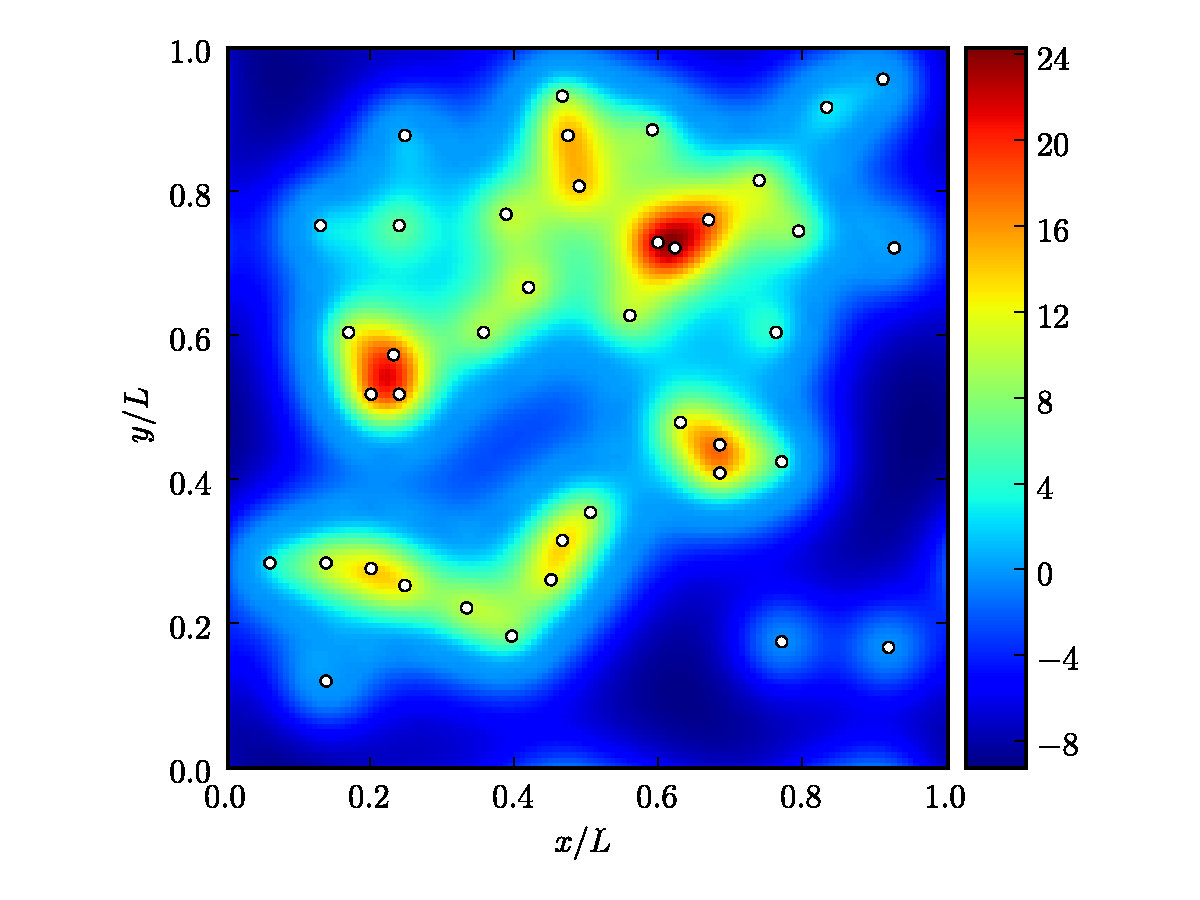}
\end{tabular}
\end{center}
\caption{(Color online)  Same as in Figs.~\ref{fig:two} and~\ref{fig:four}
but for a different distribution of impurities, $k_{\rm c}=(2\pi/L) \times 15$, and $d/L=0.07$.
\label{fig:six}}
\end{figure}

\begin{figure}
\begin{center}
\includegraphics[width=1.00\linewidth]{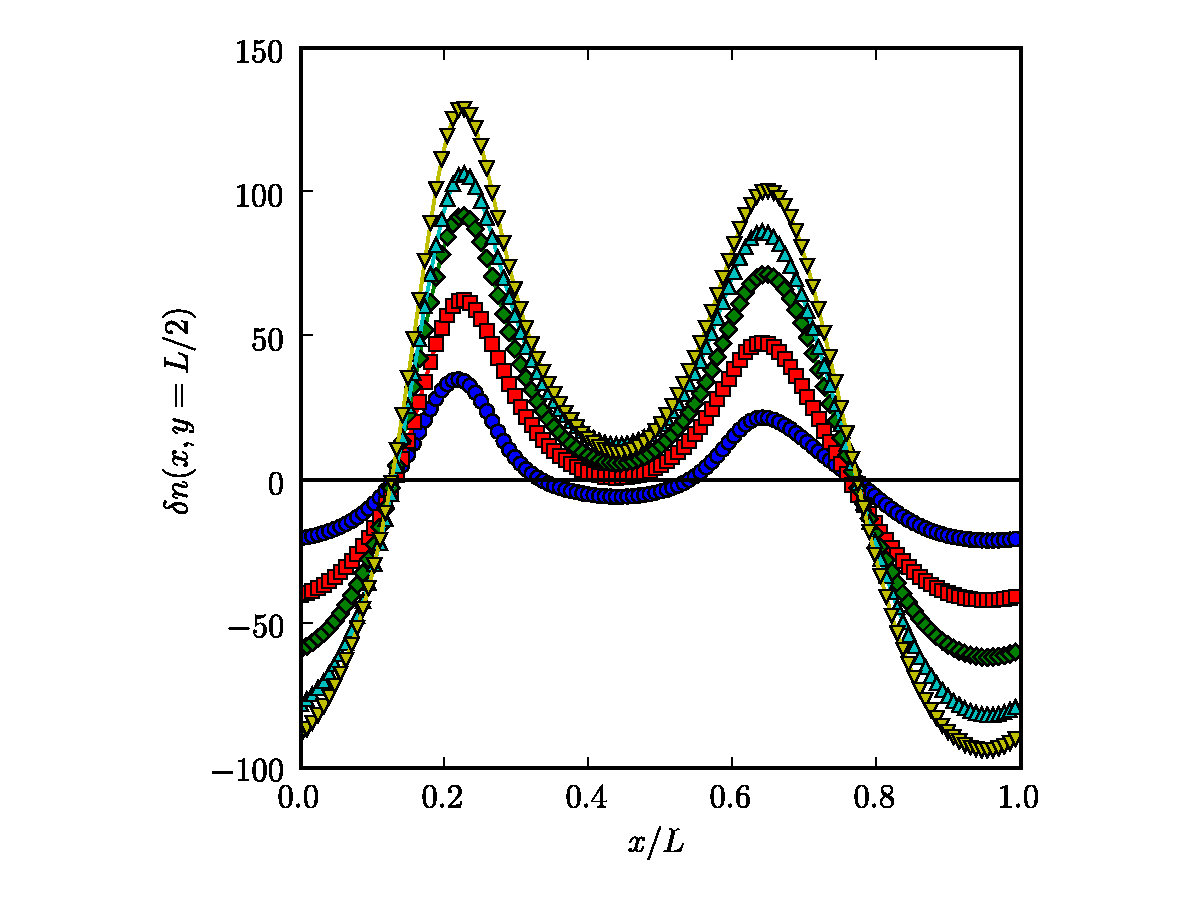}
\includegraphics[width=1.00\linewidth]{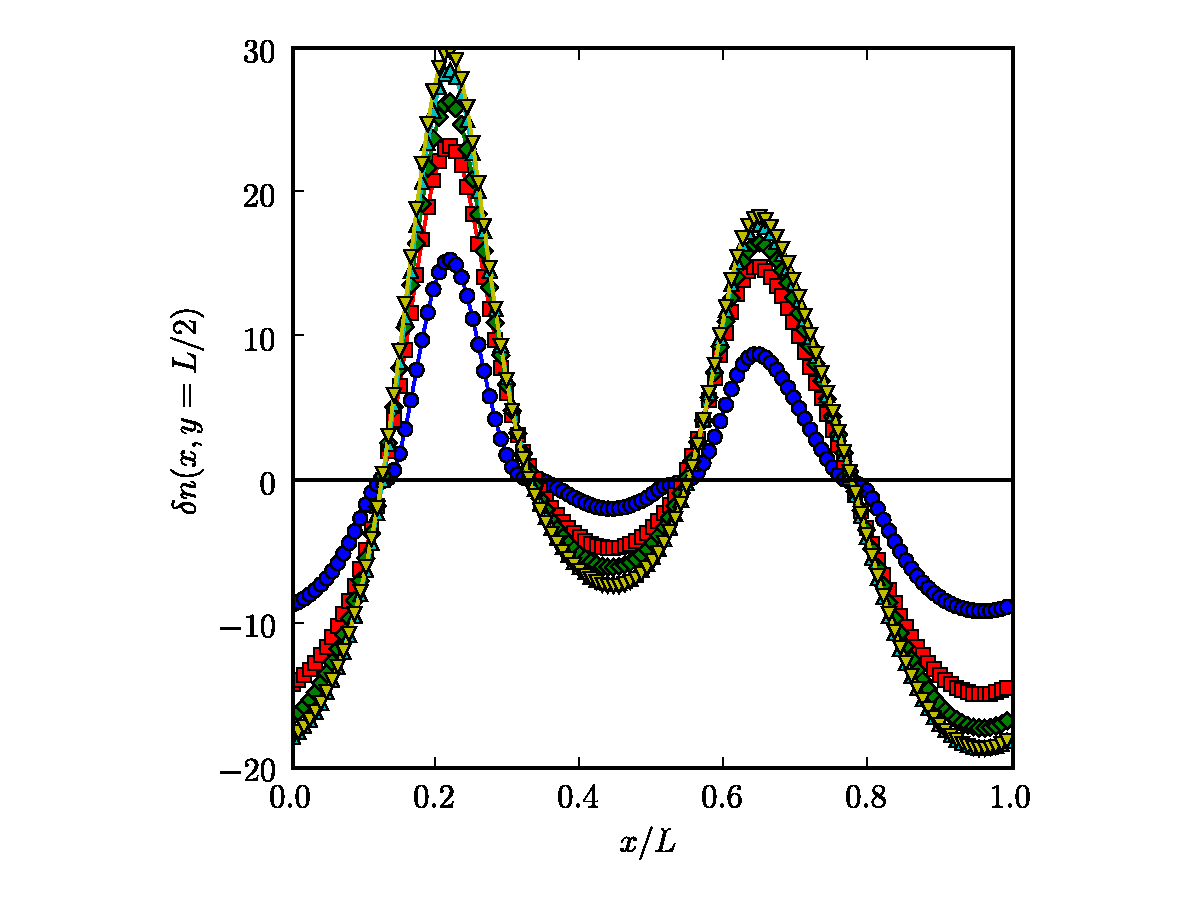}
\caption{(Color online) Illustrating the ${\cal Q}$-dependence of $\delta n ({\bm r})$. 
Top panel: A one-dimensional plot of the non-interacting density profile $\delta n({\bm r})$ 
corresponding to the external potential in the top left panel of Fig.~\ref{fig:six} as a function of $x/L$ for $y/L=0.5$. The (blue) circles label the result for ${\cal Q}=0$, the (red) squares label the result for ${\cal Q}=10$, the (green) diamonds label the result for ${\cal Q}=20$, the (cyan) triangles up label the result for ${\cal Q}=30$, and the (yellow) triangles down label the result for ${\cal Q}=40$. Bottom panel: same as in the top panel but for the full self-consistent density profile.\label{fig:seven}}
\end{center}
\end{figure}

Finally in Figs.~\ref{fig:six}-\ref{fig:seven} 
we illustrate the dependence of $\delta n({\bm r})$ on ${\cal Q}$, {\it i.e.} on a gate potentials which move
the average density-away from the Dirac-point. 
Because of the unavoidable presence of external charges in any graphene sheet environment,
this is actually the generic case.  Special neutral sheet properties, like those referred to below
in the single impurity case, will be difficult to realize experimentally.
Fig.~\ref{fig:six} shows the external potential created by a particular distribution 
of $N_{\rm imp}=40$ random charges, different again from the distributions used in producing Figs.~\ref{fig:two}-\ref{fig:four}, 
and the corresponding ground-state density profile $\delta n({\bm r})$ calculated for ${\cal Q}=0$. 
The data in Fig.~\ref{fig:six} refer to a system with square size $L^2=1922~{\cal A}_0\sim 100~{\rm nm}^2$.
We then calculate $\delta n({\bm r})$ for the {\it same} distribution of impurities but for ${\cal Q}=10,20,30$ and $40$. 
The results of these simulations are shown and compared in Fig.~\ref{fig:seven}. 
From this figure we clearly see that increasing the average density of the system, increases the amplitude of the 
density fluctuations substantially when electron-electron interactions are neglected 
(see top panel in Fig.~\ref{fig:seven}). 
When electron-electron interactions are included (see bottom panel in Fig.~\ref{fig:seven}),
this effect still occurs but $\delta n({\bm r})$ seems to saturate with increasing ${\cal Q}$.  
Of course, the carrier density fluctuation decreases in a relative sense with increasing ${\cal Q}$.

We conclude this section, by reporting results for the single-impurity case.
The calculation of the density distribution of 2D non-interacting massless Dirac fermions
in the presence of a single Coulombic impurity placed at the origin ${\bm R}_i=0$ 
of the graphene plane ($d=0$) has recently received a great deal of attention~\cite{divincenzo_prb_1986,Kolezhuk_prb_2006,Katnelson_prb_2006,Biswas_prb_2007,Shytov_prl_2007,Pereira_prl_2007,terekhov_prl_2008}. The analytical analyzes reported in these works 
shows the existence of (at least) two different regimes: (i) a regime termed ``subcritical", for $Z\alpha_{\rm ee}<1/2$, in 
which the screening density $\delta n({\bm r})$ is localized on the impurity, $\delta n ({\bm r}) \propto \delta({\bm r})$ 
and (ii) a regime termed ``supercritical", for $Z\alpha_{\rm ee}>1/2$, in which the screening density exhibits a power-law tail 
$\delta n({\bm r}) \sim 1/r^2$ at large distances. 
It is important to understand how these results are altered by the 
electron-electron interactions present in real graphene planes. 
The situation in graphene sheets is in 
this sense very different from standard semiconductor shallow-impurity problems, especially so when the 
Fermi level lies at the Dirac point. In the standard problem, it is a good approximation to 
truncate the Hamiltonian to a single band.  Interactions then play no role since, 
a single-hole or single-electron trapped by a charged impurity does not interact with itself.
In graphene, on the ohter hand, both conduction and valence bands must be retained and the 
single-impurity problem is really a many-body problem.  

\begin{figure}
\begin{center}
\includegraphics[width=1.00\linewidth]{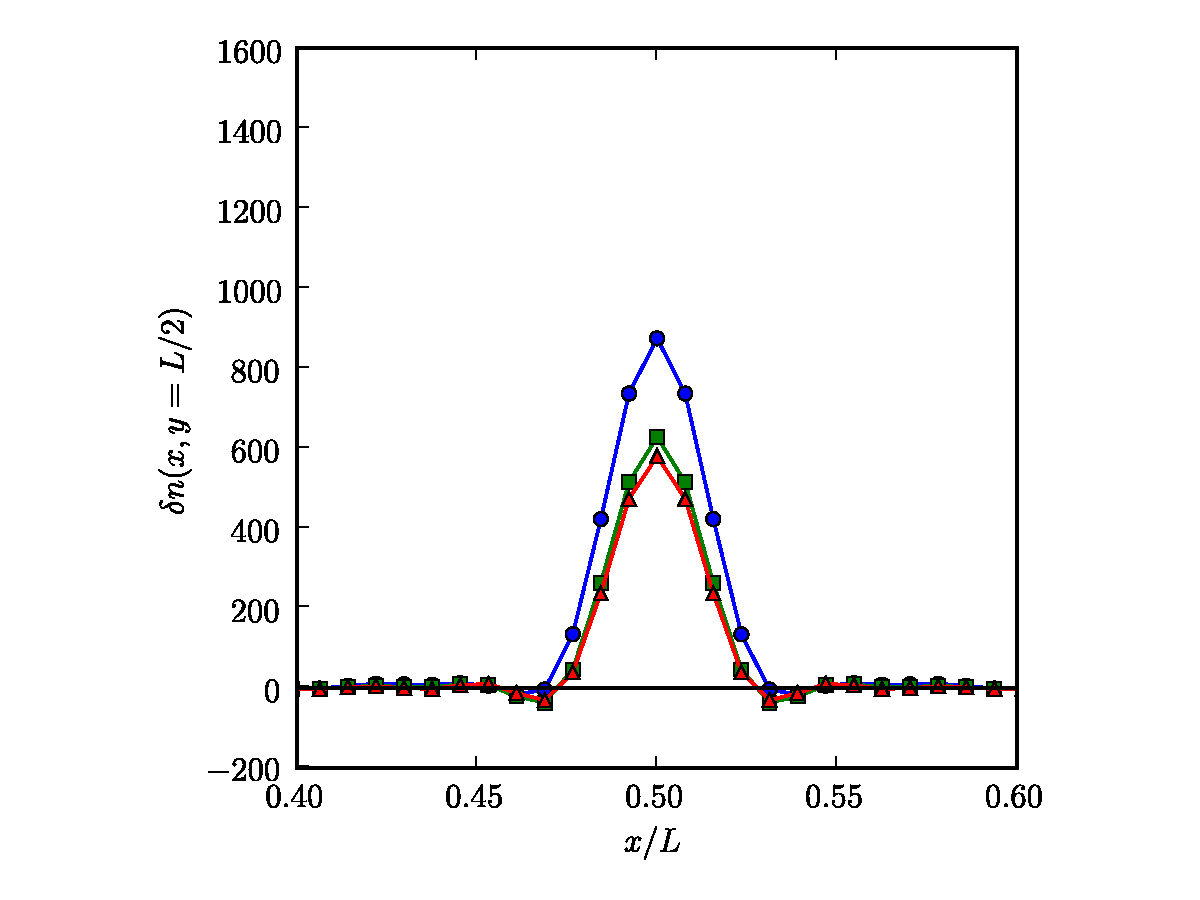}
\includegraphics[width=1.00\linewidth]{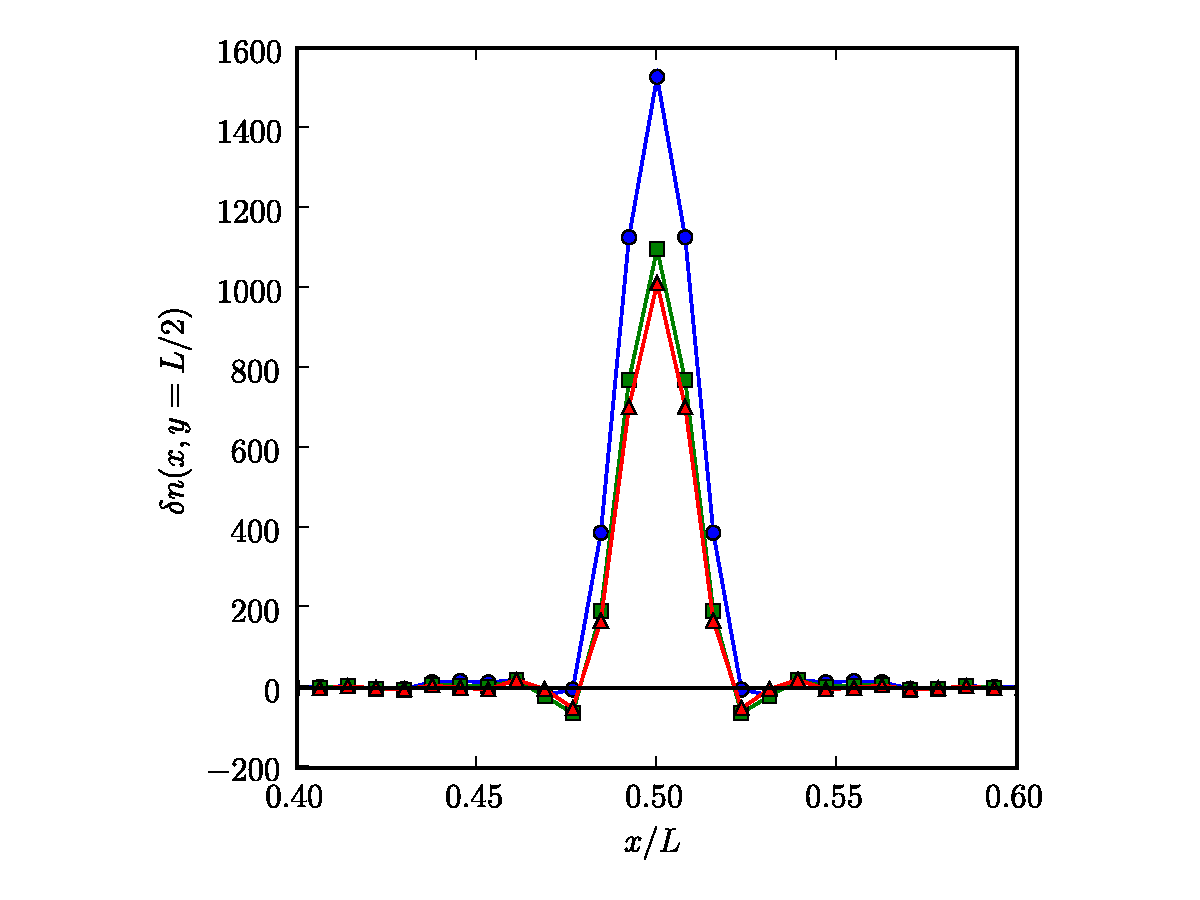}
\end{center}
\caption{(Color online). One-dimensional plots of $\delta n ({\bm r})$ as a function of $x/L$ for $y/L=0.5$ for
a single impurity with $Z=+1$ located at $x=y=L/2$. Here $d=0.0$ and $\alpha_{\rm ee}=0.5$. 
Top panel: numerical results for $k_{\rm c}=(2\pi/L) \times 15$. Bottom panel: numerical results for $k_{\rm c}= (2\pi/L) \times 20$. The (blue) circles label the non-interacting result, the (green) squares label the self-consistent Hartree-only result, 
and the (red) triangles label the full self-consistent result. \label{fig:eight}}
\end{figure}

The method used here to solve  
the Kohn-Sham-Dirac equations, in which we project onto a plane-wave basis, 
is not optimized for the study the single impurity problem because it 
does not take advantage of its circular symmetry. 
Nonetheless, in Fig.~\ref{fig:eight} we present some numerical results 
for the density distribution of a 2D CEG in the presence of 
a single impurity placed at the center of the sample $(x=L/2,y=L/2)$ 
and right on the graphene plane. In particular, we show a 1D plot of $\delta n ({\bm r})$ as a function of $x/L$ for $y/L=0.5$. 
These density profiles correspond to a $Z=+1$ impurity in a Dirac sea with $\alpha_{\rm ee}=0.5$ and ${\cal Q}=0$. 
In the two simulation results reported in this figure we have used $k_{\rm c}=(2\pi/L)\times 15$, which 
corresponds to an effective square size of 
$L^2=1922~{\cal A}_0\sim 100~{\rm nm}^2$ 
and $k_{\rm c}=(2\pi/L)\times 20$, which corresponds to an effective square size of 
$L^2=3362~{\cal A}_0\sim 175~{\rm nm}^2$. 
Comparing the results in the top [$k_{\rm c}=(2\pi/L)\times 15$] and bottom [$k_{\rm c}=(2\pi/L)\times 20$] 
panels we can clearly see how they are compatible with a completely 
localized screening density with a $\delta$-function shape, 
the finite-width of $\delta n({\bm r})$ being solely due to our momentum-space cutoff. 

\begin{figure}
\begin{center}
\includegraphics[width=1.00\linewidth]{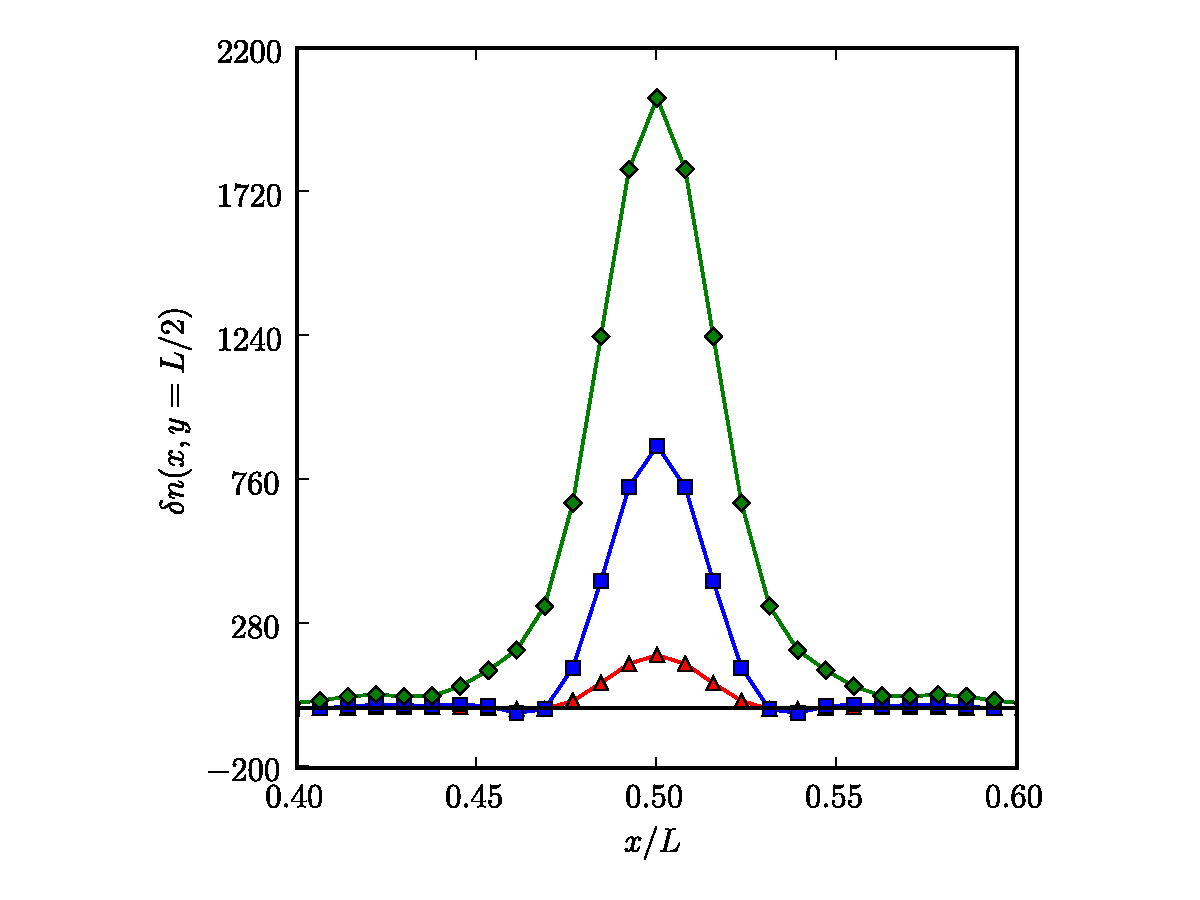}
\includegraphics[width=1.00\linewidth]{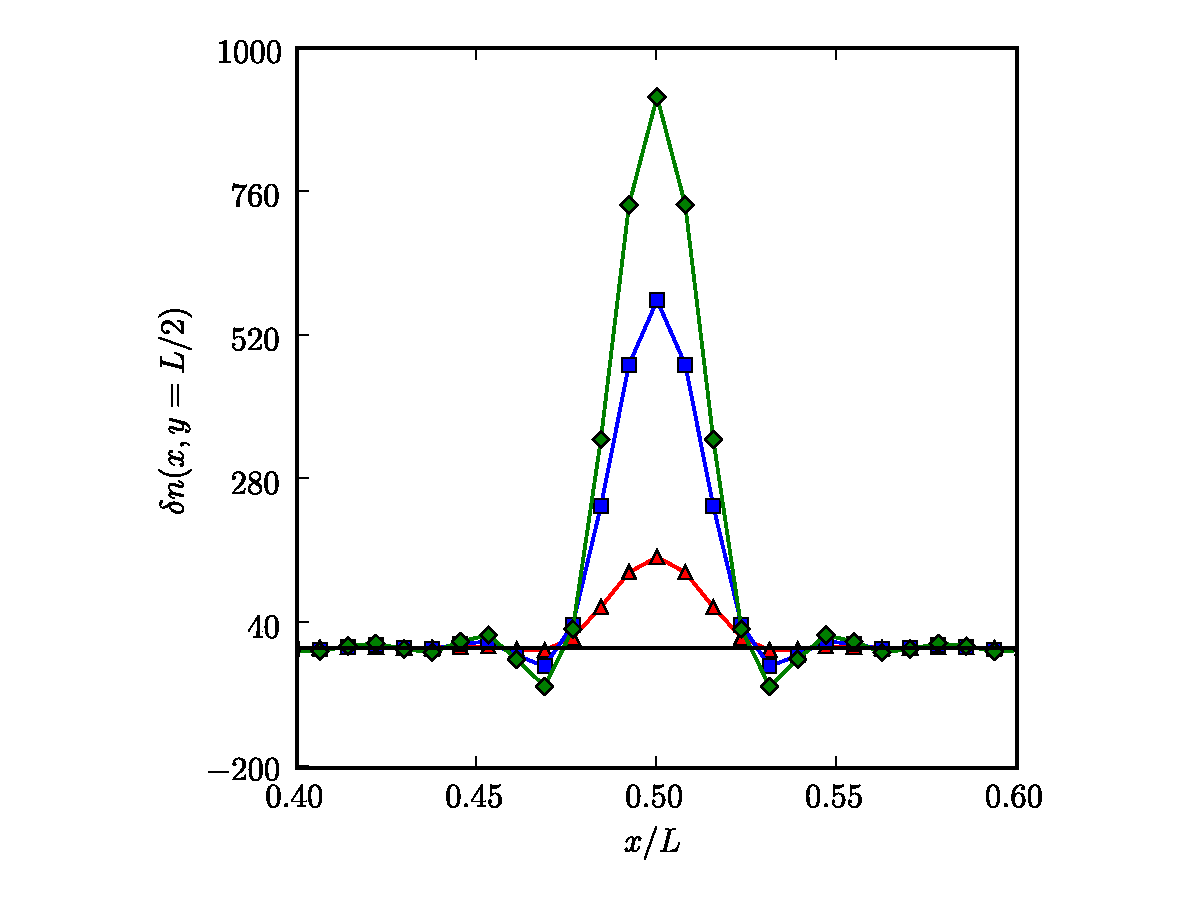}
\end{center}
\caption{(Color online). One-dimensional plots of $\delta n ({\bm r})$ as a function of $x/L$ for $y/L=0.5$ for
one impurity with $Z=+1$ located at $x=y=L/2$. Here $d=0.0$ and $k_{\rm c}=(2\pi/L)\times 15$. 
Top panel: non-interacting results. Bottom panel: full results. 
In each panel the (red) triangles label the results for $\alpha_{\rm ee}=0.1$, 
the (blue) squares label the results for $\alpha_{\rm ee}=0.5$, while
the (green) diamonds label the results for $\alpha_{\rm ee}=1.0$.
\label{fig:nine}}
\end{figure}

Finally, in Fig.~\ref{fig:nine} we show how $\delta n({\bm r})$ behaves quite differently in the two cases 
$\alpha_{\rm ee}=0.1$ and $\alpha_{\rm ee}=1.0$. Indeed, the non-interacting density seems to possess a long-range tail for $\alpha_{\rm ee}=1.0$. When electron-electron interactions are taken into account though, it seems that the behavior of $\delta n({\bm r})$ is quite similar in both cases. This is in agreement with the findings of Ref.~\onlinecite{terekhov_prl_2008}, in which the authors have shown that when electron-electron interactions are taken into account 
at the Hartree level, a $Z=+1$ impurity always remains in the subcritical regime.

\section{Summary and Discussion}
\label{sect:conclusions}

When inter-valley scattering is weak, doped and gated graphene sheets can be described using an 
envelope-function Hamiltonian with a new sublattice pseudospin degree-of-freedom,
an ultrarelativistic massless-Dirac free-fermion term, a pseudospin scalar disorder potential, 
and a non-relativistic instantaneous Coulombic interaction term.  
There is considerable evidence from
experiment that this simplified description of a honeycomb lattice of 
Carbon atoms is usually a valid starting point for theories of those observables that depend
solely on the electronic properties of $\pi$-electrons near the graphene Dirac point.  
Although the use of this model simplifies the physics considerably it still leaves us with
a many-body problem without translational invariance which we do not know how to solve.

A common strategy in piecing together the physics of disordered interacting-fermion problems 
is to solve models in which interactions are neglected, appealing perhaps to Fermi-liquid-theory 
concepts, and to solve problems in which disorder is neglected, hoping that it is sufficiently 
weak to be unimportant for some observations.  We anticipate that this {\em divide and conquer}
approach will often fail in graphene.  With this motivation, we have presented in this paper 
a Dirac-Kohn-Sham density-functional-theory scheme for graphene sheets, which treats interactions 
and smooth inhomogeneous external potentials on an equal footing. Although it is formally an exact solution of the graphene 
many-body problem, it relies in practice on approximate exchange-correlation functionals.

The best approximation available for the graphene problem at present is the local-density-approximation (LDA) 
for the exchange-correlation potential.
In this paper we have provided convenient parametrizations of the exchange and correlation energies 
of uniform-density graphene systems based on random-phase-approximation many-body calculations.
These results can be used to take account not only of density variations in a 
disordered graphene sheet, but also of changes in the sheets dielectric environment which 
alters the coupling constant which appears in the Dirac model for graphene.  

We believe that the exchange and correlation effects captured by our DFT theory will be 
important for many-qualitative aspects of grahene electronic structure.  In graphene the 
dependence of the LDA exchange-correlation potential on density is opposite to that of 
normal 2D or 3D electron systems.  As explained in detail in Ref.~\onlinecite{ourdgastheory},
the origin of this behavior is in the interplay between Dirac-model free-fermion pseudospin-chirality
and Coulomb interactions; when the carrier density is zero in a graphene sheet the 
pseudospin-chirality polarization is maximized and this leads to lower interaction energies.

It is important to contrast the DFT scheme outlined in this paper with 
normal microscopic DFT applied to the carbon atoms of a graphene sheet.  
The fully microscopic DFT deals with all the Carbon atom orbitals, 
including the $sp^{2}$ bonding and anti-bonding orbitals, which are 
away from the Fermi level and neglected in the Dirac-model, and can be used 
for example~\cite{phonons} to calculate the electron-phonon coupling 
in a graphene sheet from first principles.  
Microscopic DFT also provides an {\em ab initio} 
estimate of the massless Dirac velocity, which is a 
phenomenological parameter of the Dirac-model theory.
The advantages of using the present DFT scheme for some 
$\pi$-orbital properties of graphene sheets are made clear by 
observing that microscopic DFT, in which the exchange-correlation 
potential is based on the properties of a uniform 3D 
electron gas, fails to capture the anomalous sign of the density-derivative of 
graphene's exchange-correlation potential. 
From a microscopic point of view this anomalous sign is a 
combined consequence of the peculiarities of Dirac bands and 
non-local exchange and correlation effects captured by the 
uniform-density Dirac-model.  

In this paper we have illustrated the properties of this DFT description of 
disordered graphene sheets by concentrating on the non-uniform carrier density.
Although the Kohn-Sham orbitals which appear in this and other DFT scheme are 
formally justified {\it only} for the role they play in density and ground-state-energy 
calculations (due to the Hohenberg-Kohn theorem~\cite{dft}), 
their physical significance is often interpreted more liberally by
associating the Kohn-Sham eigenvalues with quasiparticle 
energies.  This pragmatic approach can fail spectacularly, as it famously does 
in the estimation of common semiconductor band gaps, but is more often quite useful 
in interpreting spectral properties of materials.  
In the case of $\pi$-orbital properties of disordered graphene sheets,
STM local-density-of-states, ARPES, and optical conductivity spectra require 
interpretation.  In our view it will be useful to apply the present approach as 
one element of an effort to improve understanding of what these probes tell us about 
particular graphene sheets.  The fact that the self-energy of uniform-density 
graphene sheets has a large dependence on wavevector relative to the Dirac-point~\cite{ourdgastheory},
in addition to its dependence on wavevector and energy relative to the Fermi surface, 
may help justify taking this liberty with the DFT formalism.

\acknowledgments

We gratefully acknowledge useful discussions 
with Victor Brar, Misha Fogler, Kentaro Nomura, and Gaetano Senatore. 
M.P. acknowledges the kind hospitality of the Department of Physics of the University of Texas 
at Austin during the final stages of the preparation of this work.
A.H.M. was supported by the Welch Foundation, the ARO, and SWAN-NRI.
The figures have been prepared with the Open Source scipy/numpy/matplotlib 
packages of the Python programming language.

\end{document}